\documentclass[prd, nofootinbib, notitlepage, twocolumn, superscriptaddress]{revtex4-1}

\usepackage[caption=false]{subfig}
\usepackage{graphicx}
\usepackage{amsmath}

\usepackage{xcolor}
\usepackage{ulem}
\definecolor{naviBlue}{RGB}{0,0,128}
\usepackage[colorlinks  = true,
            linkcolor   = naviBlue,
            urlcolor    = naviBlue,
            citecolor   = naviBlue,
            anchorcolor = naviBlue,
            backref     = page       ]{hyperref}
            
\newcommand{\gsim}{\lower.7ex\hbox{$\;\stackrel{\textstyle>}{\sim}\;$}}
\newcommand{\lsim}{\lower.7ex\hbox{$\;\stackrel{\textstyle<}{\sim}\;$}}            

\renewcommand*{\backref}[1]{}
\renewcommand*{\backrefalt}[4]{
  \small
    \ifcase #1 
    \or        (Cited on page~#2.)
    \else      (Cited on pages~#2.)
    \fi
}

\newcommand{\appref} [1]{\hyperref[app::#1]{Appendix~\ref*{app::#1}}}
\newcommand{\secref} [1]{\hyperref[sec::#1]{Sec.~\ref*{sec::#1}}}
\newcommand{\figref} [1]{\hyperref[fig::#1]{Fig.~\ref*{fig::#1}}}
\newcommand{\tabref} [1]{\hyperref[tab::#1]{Table~\ref*{tab::#1}}}
\newcommand{\eqnref} [1]{\hyperref[eqn::#1]{Eq.~(\ref*{eqn::#1})}}
\newcommand{\noteref}[1]{\textsuperscript{\ref{note::#1}}}

\definecolor{Sketch}{HTML}{965454}

\usepackage{lipsum}
\usepackage{lineno}

\newcommand{\Tp}{T_p}
\newcommand{\Tproj}{T_i}
\newcommand{\pb}{{\bar{p}}}
\newcommand{\nb}{{\bar{n}}}
\newcommand{\Tpbar}{T_{\bar{p}}}
\newcommand{\sS}{\sqrt{s}}
\newcommand{\xR}{x_\mathrm{R}}
\newcommand{\pT}{p_{\mathrm{T}}}
\newcommand{\pL}{p_{\mathrm{L}}}
\newcommand{\cosT}{\cos(\theta)}

\newcommand{\sigmaInv}{\sigma_{\mathrm{inv}}}
\newcommand{\sigmaRel}{\sigma^{\mathrm{rel}}}

\begin{document}

\title{Prescriptions on antiproton cross section data for precise theoretical antiproton flux predictions}

\author{Fiorenza Donato}
\email{donato@to.infn.it}
\affiliation{Dipartimento di Fisica, Universit\`a di Torino, Via P. Giuria 1, 10125 Torino, Italy}
\affiliation{Istituto Nazionale di Fisica Nucleare, Sezione di Torino, Via P. Giuria 1, 10125 Torino, Italy}
\author{Michael Korsmeier}
\email{korsmeier@physik.rwth-aachen.de}
\affiliation{Dipartimento di Fisica, Universit\`a di Torino, via P. Giuria 1, 10125 Torino, Italy}
\affiliation{Istituto Nazionale di Fisica Nucleare, Sezione di Torino, Via P. Giuria 1, 10125 Torino, Italy}
\affiliation{Institute for Theoretical Particle Physics and Cosmology, RWTH Aachen University, 52056 Aachen, Germany}
\author{Mattia Di Mauro}
\email{mdimauro@slac.stanford.edu}
\affiliation{W.W. Hansen Experimental Physics Laboratory, Kavli Institute for Particle Astrophysics and Cosmology, Department of Physics and SLAC National Accelerator Laboratory, Stanford University, Stanford, California 94305, USA}


\begin{abstract}
\noindent
After the breakthrough from the satellite-borne PAMELA detector, the flux of cosmic-ray (CR) antiprotons has been provided with unprecedented accuracy by AMS-02 on the International Space Station. Its data
spans an energy range from below 1 GeV up to 400 GeV and most of the data points contain errors below the amazing level of 5\%. 
\\
The bulk of  the antiproton flux is expected to be produced by the scatterings of CR protons and helium off interstellar hydrogen and helium atoms at rest.
The modeling of these interactions, which requires the relevant production cross sections, induces an uncertainty in the determination of the antiproton source term that can  even exceed the uncertainties in the CR $\pb$ data itself.  
\\
The aim of the present analysis is to determine the uncertainty required for $p+p\rightarrow \pb + X$ cross section measurements such that the induced uncertainties on the $\pb$ flux are at the same level.
Our results are discussed both in the center-of-mass reference frame, suitable for collider experiments, and in the laboratory frame, as occurring in the Galaxy. 
We find that 
cross section data should be collected with accuracy better that few percent with proton beams from 10 GeV to 6 TeV and a pseudorapidity $\eta$ ranging from 2 to almost 8 or, alternatively,  
with  $\pT$ from 0.04 to 2 GeV and $\xR$ from 0.02 to 0.7. Similar considerations hold for the $p$He production channel. 
The present collection of data is far from these requirements. Nevertheless,  they could, in principle, be reached by fixed target experiments with beam energies in the  reach of CERN accelerators. 
\end{abstract}

\maketitle

\section*{\label{sec::introduction}Introduction}

Astroparticle physics of charged cosmic rays (CRs) has become a high-precision discipline in the last decade. 
Spaceborne experiments like first PAMELA and more recently and still operating AMS-02 have reduced the measurement uncertainties of CR fluxes to the percent level over an energy range from below 1 GeV up to a few TeV, which is typical for Galactic CRs. 
In particular, this result has been achieved for the nuclear  
\cite{PAMELA_Adriani:2011cu,2014ApJ...791...93A,AMS-02_Aguilar:2015_ProtonFlux,AMS-02_Aguilar:2015_HeliumFlux,Aguilar:2016vqr} and leptonic (positron and electron)  
\cite{2009Natur.458..607A,2011PhRvL.106t1101A,2013PhRvL.110n1102A,2014PhRvL.113l1102A,2014PhRvL.113v1102A} components, and also for the CR antiprotons \cite{2010PhRvL.105l1101A,AMS-02_Aguilar:2016_AntiprotonFlux}. 
The rare CR antimatter has been extensively studied as a possible indirect signature of dark matter annihilating in the halo of the Galaxy 
\cite{2010pdmo.book..521S} (and Refs. therein). 
The recent AMS-02 data on the antiproton flux and $\pb/p$ flux ratio has reached an unprecedented precision from about 1 GeV up to  hundreds of GeV
\cite{AMS-02_Aguilar:2016_AntiprotonFlux}. 
\\
This exceptional experimental accuracy poses the challenging task of a theoretical interpretation with an uncertainty at a similar level. 
The dominant part of the  antiprotons in our Galaxy arises from secondary production, namely it originates by the inelastic scattering of incoming CRs off 
interstellar medium (ISM) nuclei at rest. 
In practice, the secondary antiprotons are produced from the scattering of CR $p$ and He on ISM consisting again of H and He.
The number of produced antiprotons then depends on the correct modeling of the production cross section $d\sigma (p+p \rightarrow \pb + X)/dT_\pb$ and the equivalent reactions with He instead of $p$. 
The production cross sections induce a non-negligible uncertainty in the prediction of the secondary antiprotons, as already underlined in 
\cite{Donato:2001_DTUNUC,Bringmann:2006im,Donato:2008jk}. 
In literature, there have been different approaches to describe the $\pb$ production cross section; after the first parametrization 
for the $pp$ scattering \cite{TanNg:1983_pbarCrossSectionparametrization}, which is now probably outdated, 
Monte Carlo (MC) predictions have been employed, in particular, for the He channels using the DTUNUC code \cite{Simon_Antiproton_CS_Scaling_1998,Donato:2001_DTUNUC}. For these He channels measurements were not available until
very recently when the LHCb Collaboration presented a preliminary analysis on the search for antiprotons in collisions of 6.5~TeV protons on a fixed helium target at the LHC \cite{LHCb_pHe}. 
A parametrization deduced from a large $pp$ and $pA$ (proton-nuclei) data  set was proposed in \cite{Duperray:2003_pbarCrossSectionparametrizationForPA}. 
Recently, triggered by the NA49 data \cite{NA49_Anticic:2010_ppCollision}, new parametrizations have been proposed in 
\cite{diMauro:2014_pbarCrossSectionparametrization,Kappl:2014_pbarCrossSection} as well as predictions from MC  generators tuned with LHC data 
\cite{Kachelriess:2015_pbarCrossSection}. 
Nevertheless, the theoretical uncertainty induced by the modeling of fundamental interactions on the  antiproton spectrum is sizable, reaching a few ten percent. 
\\
In this paper we  ``backwards engineer'' the usual process of cross section parametrization in order to determine 
the accuracy required on cross section measurements to match AMS-02 accuracy. 
Our aim is to provide, for the first time, quantitative indications for future high-energy experiments about the kinematical regions and the precision level to be obtained, in order to induce uncertainties on $\pb$ flux which do not exceed the uncertainty in the present CR data. 
\\
This paper is structured as follows. In \secref{theory} we review the main steps for the calculation of the antiproton source term starting from the invariant cross section. In \secref{methods} we explain how we invert this calculation in order to assign uncertainty requirements on the differential cross section. The results are presented in \secref{results} and are summarized in \secref{conclusion}.

\section{\label{sec::theory}Theoretical framework for the cosmic antiproton source spectrum}

\begin{figure}[b!]
	\includegraphics[width=1.\linewidth]
	    {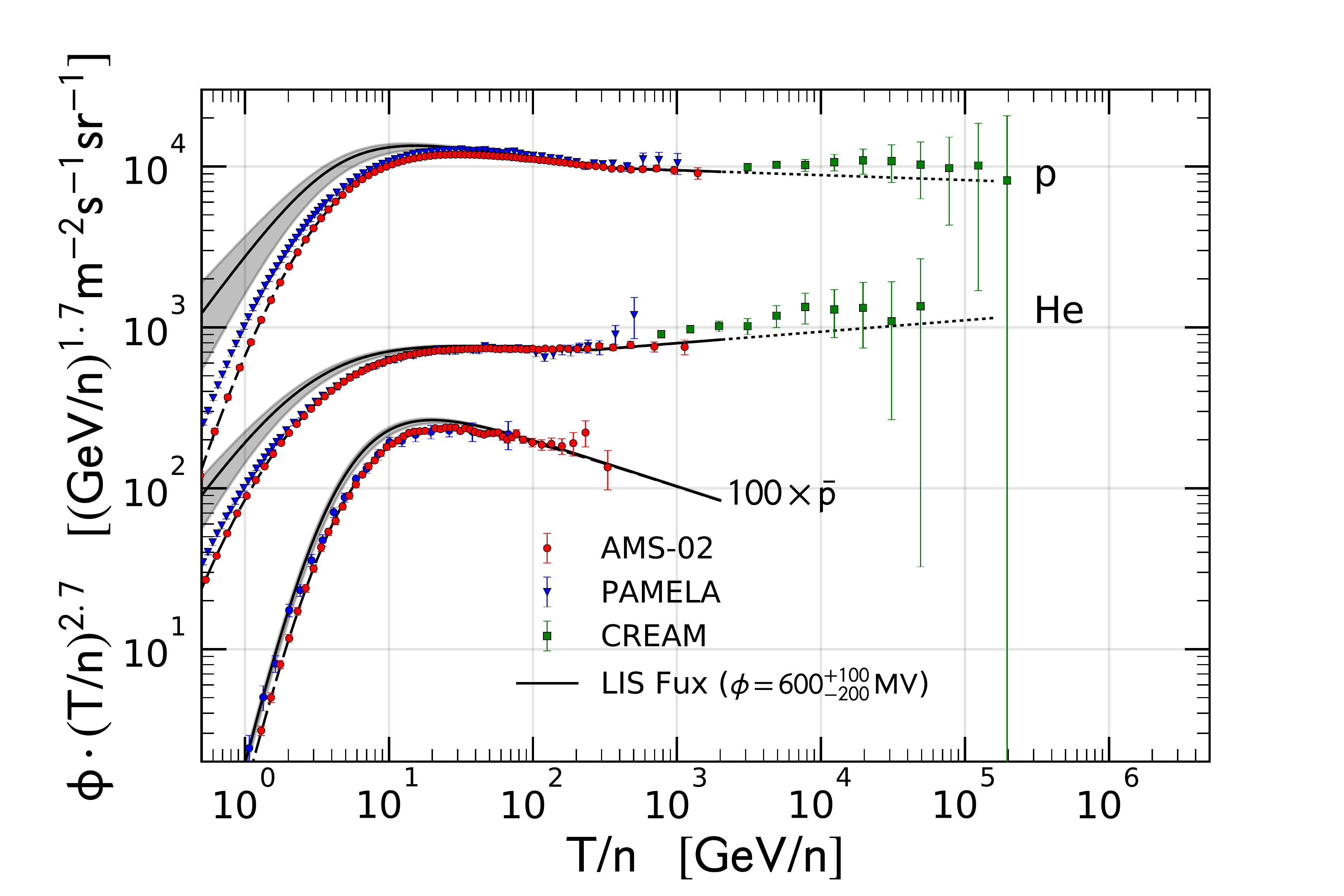}
  \caption{Recent flux measurements for CR protons, helium, and antiprotons by AMS-02 
\cite{AMS-02_Aguilar:2015_ProtonFlux, AMS-02_Aguilar:2015_HeliumFlux, AMS-02_Aguilar:2016_AntiprotonFlux}, PAMELA \cite{PAMELA_Adriani:2011cu, PAMELA_Adriani:2012paa}, 
           and CREAM \cite{CREAM_Yoon:2011aa}. 
           The energy-differential fluxes $\phi$ are given as a function of kinetic energy per nucleon $T/n$. 
           Furthermore, the IS fluxes, demodulated in the force-field approximation 
           with an modulation potential of $\phi_\odot = 600_{-200}^{+100}$ MV, are presented. }
	\label{fig::FluxMeasurements_AMS-02}
\end{figure}

Antiprotons in our Galaxy are dominantly produced in  processes of CR nuclei colliding with ISM.
Hence, the ingredients to calculate the $\pb$ source term, i.e., the number of antiprotons per volume, time, and energy, 
are the flux of the incident CR species $i$, $\phi_i$, and the density of the ISM component $j$, where, in practice, both $i$ and $j$ are $p$ and He.
The source term for secondary antiprotons is given by a convolution integral of the CR flux, the ISM targets, and the relevant cross section:
\begin{eqnarray}
\label{eqn::sourceTerm_1}
  q_{ij}(\Tpbar) &=& \int\limits_{T_{\rm th}}^\infty d\Tproj \,\, 4\pi \,n_{\mathrm{ISM},j} \, \phi_i  (\Tproj) \, \frac{d\sigma_{ij}}{d \Tpbar}(\Tproj , \Tpbar).
\end{eqnarray}
Here $n_{\mathrm{ISM}}$ is the ISM density and $T_{\rm th}$ the production energy threshold. 
The factor $4\pi$ corresponds to the already executed angular integration of the isotropic flux $\phi$.
The according fluxes are known precisely at the top of the Earth's atmosphere (TOA) due to AMS-02 measurements \cite{AMS-02_Aguilar:2015_ProtonFlux, AMS-02_Aguilar:2015_HeliumFlux} presented in \figref{FluxMeasurements_AMS-02}, together with the results from the precursor satellite-borne PAMELA experiment \cite{PAMELA_Adriani:2011cu, PAMELA_Adriani:2012paa} and the data from the balloon-borne CREAM detector at higher energies \cite{CREAM_Yoon:2011aa}.
At low energies $E\lsim $ 20 GeV/nucleon (in the following GeV/n) the charged particles arriving at the Earth are strongly affected by solar winds, commonly referred to as solar modulation \cite{Parker:1958_SolarModulation, Gleeson:1968_SolarModulation}, given their activity modulation on a cycle of roughly 11 years. 
We will work with interstellar (IS) quantities. The $p$ and He IS fluxes are inferred by demodulated AMS-02 data, which we obtain within the force-field approximation \cite{Fisk:1976_SolarModulation} assuming an average Fisk potential of $\phi_\odot=600$~MeV for the period of data taking \cite{Usoskin:2005_SolarModulation, Usoskin:2011_SolarModulation}. 
More complete studies on solar modulation 
 take into account time dependent proton flux data from PAMELA  and 
recent ISM flux measurements by VOYAGER \cite{Cholis:2016_Solar_Modulation, Ghelfi:2016pcv,Corti:2015bqi}. They find similar values for $\phi_\odot$.
The source term derivation only includes incoming proton energies $E_p >7 m_p \sim 6.6$ GeV ($E_p >4 m_p$) corresponding to the $\pb$ production threshold in $pp$ ($p$He) collisions. For these energies the solar modulation, which becomes negligible above a few 10 GeV, agrees reasonably well with the simple force-field approximation. 
The scattering sights are the ISM elements H and He with density given by 1 and 0.1 $\mathrm{cm^{-3}}$ in the Galactic disk respectively.
\begin{figure}[b!]
  \includegraphics[width=1.\linewidth]
	    {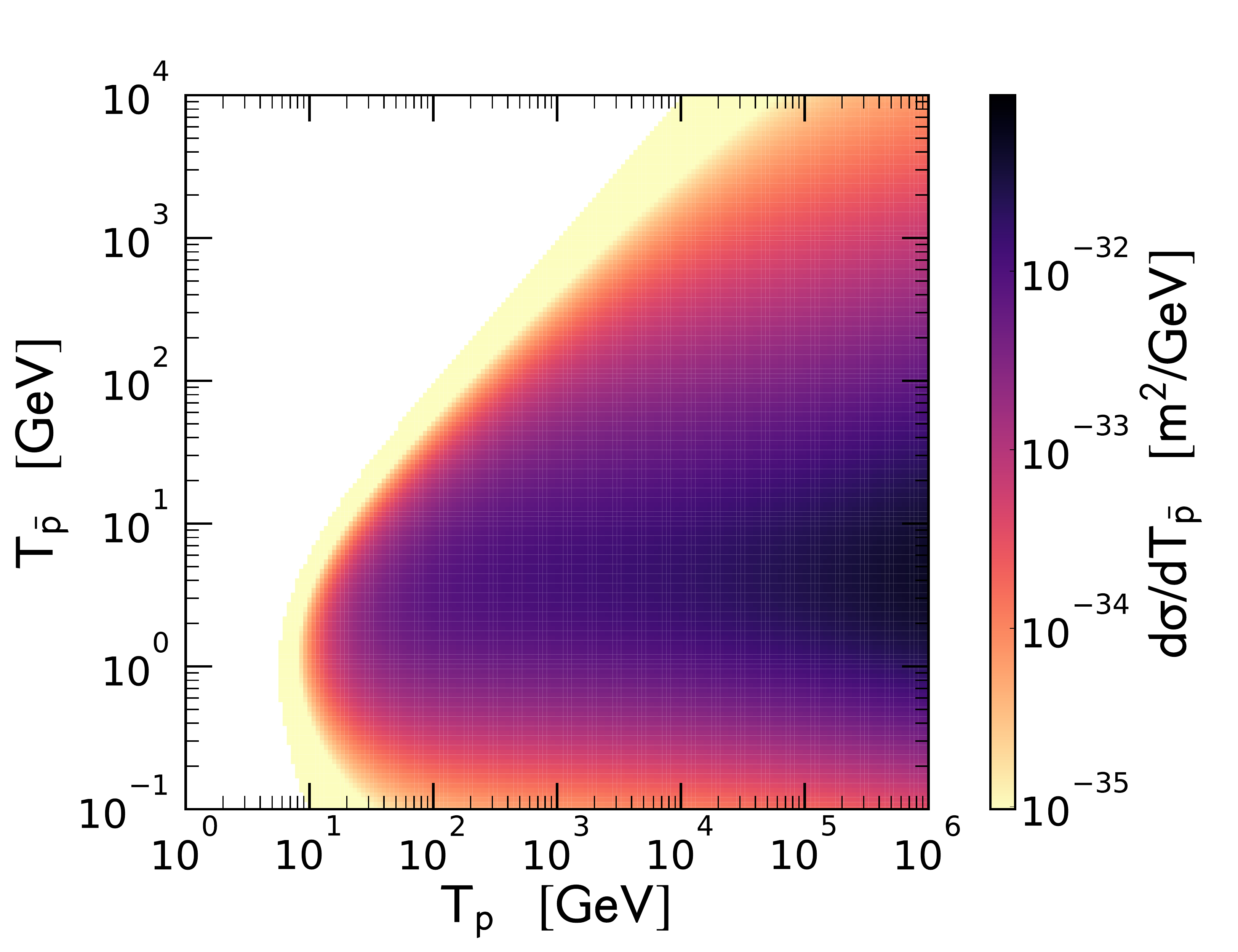}
  \caption{Energy-differential antiproton production cross section from $pp$ collisions in the LAB frame 
           as function of proton and antiproton kinetic energy $\Tp$ and $\Tpbar$, respectively. 
           The shown cross section is derived from the Di Mauro {\textit et al.} \cite{diMauro:2014_pbarCrossSectionparametrization}
           parametrization (their Eq.~12).  }
	\label{fig::EnergyDifferentialCS_LAB_diMauro}
\end{figure}

The final essential ingredient to calculate the source term 
is the cross section corresponding to the production reaction $ \mathrm{CR}_i + \mathrm{ISM}_{j} \rightarrow \bar{p} + X$
\begin{eqnarray}
\label{eqn::energyDifferentialCrossSection}
  \frac{d\sigma_{ij}}{d \Tpbar}(\Tproj , \Tpbar), 
\end{eqnarray}
where $\Tpbar$ is the kinetic energy of the produced antiproton in collisions of the CR species $i$ with kinetic energy $\Tproj$ on the 
ISM  component $j$. In the following we will call the quantity in \eqnref{energyDifferentialCrossSection} the energy-differential 
cross section\footnote{Note that $dT=dE$ and, hence, $d \sigma/dE = d\sigma/dT$.}. 
One example, derived from the cross section parametrization in Ref.  \cite{diMauro:2014_pbarCrossSectionparametrization} 
for the $pp$ channel, is shown in \figref{EnergyDifferentialCS_LAB_diMauro} as a function of $\Tpbar$ and $T_p$. The kinetic energy threshold at $\Tp=6m_p$ is clear.

\begin{figure}[t!]
	\subfloat[]{  \vspace{-1cm}\includegraphics[width=0.45\textwidth, trim={0 0.5cm 0 2cm},clip]{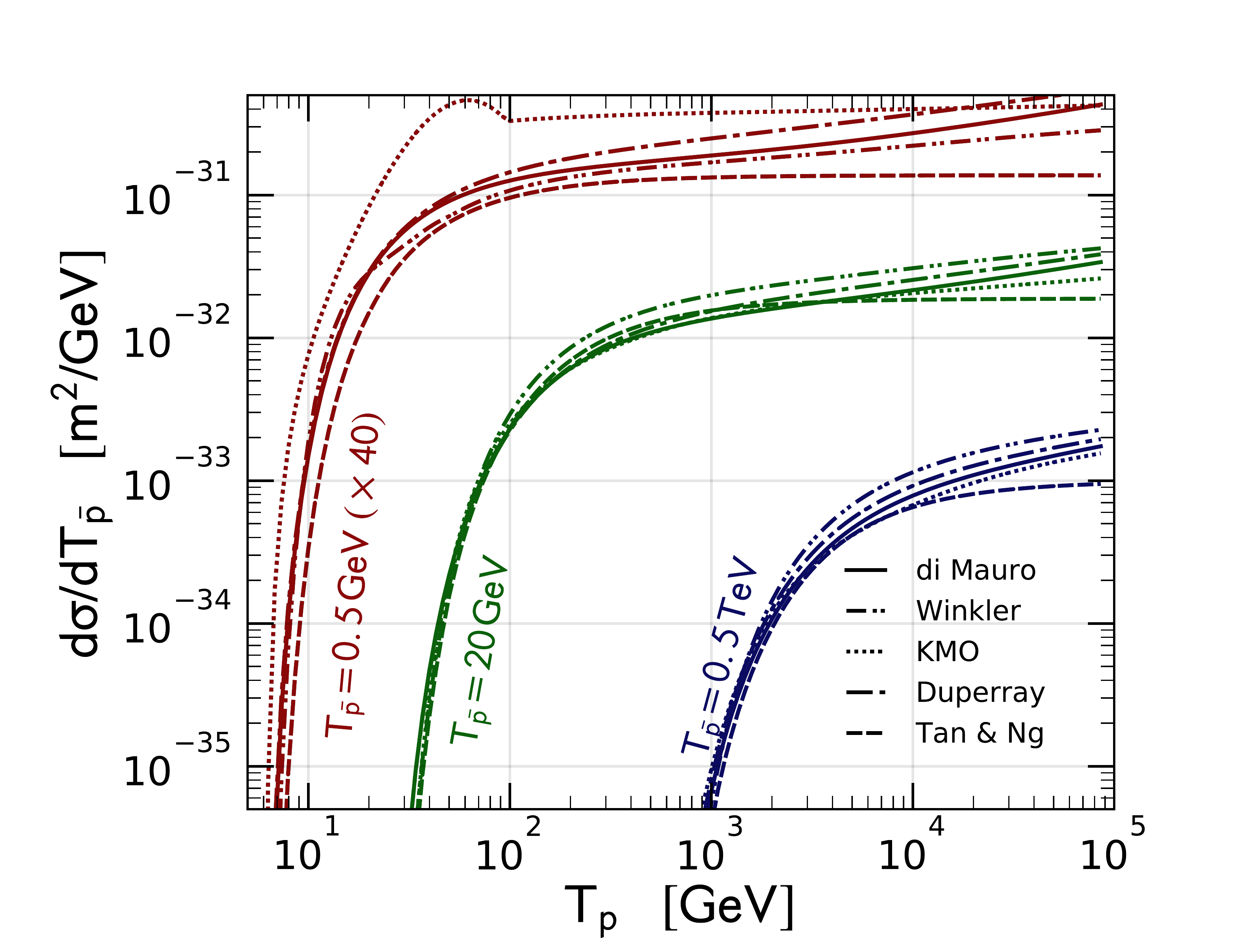}   } \\
	\subfloat[]{  \vspace{-1cm}\includegraphics[width=0.45\textwidth, trim={0 0.5cm 0 2cm},clip]{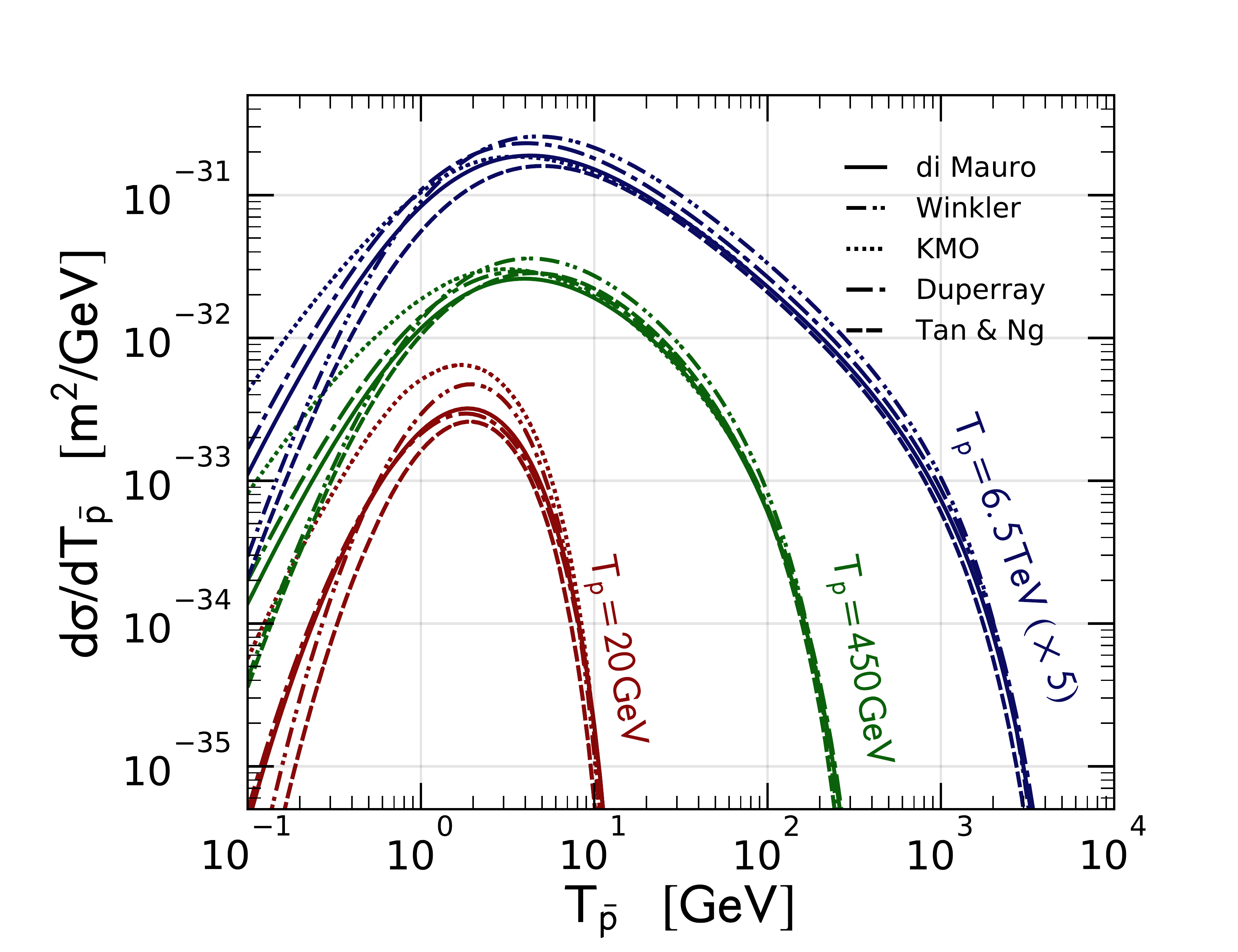}   } \\
	\subfloat[]{  \vspace{-1cm}\includegraphics[width=0.45\textwidth, trim={0 0.5cm 0 2cm},clip]{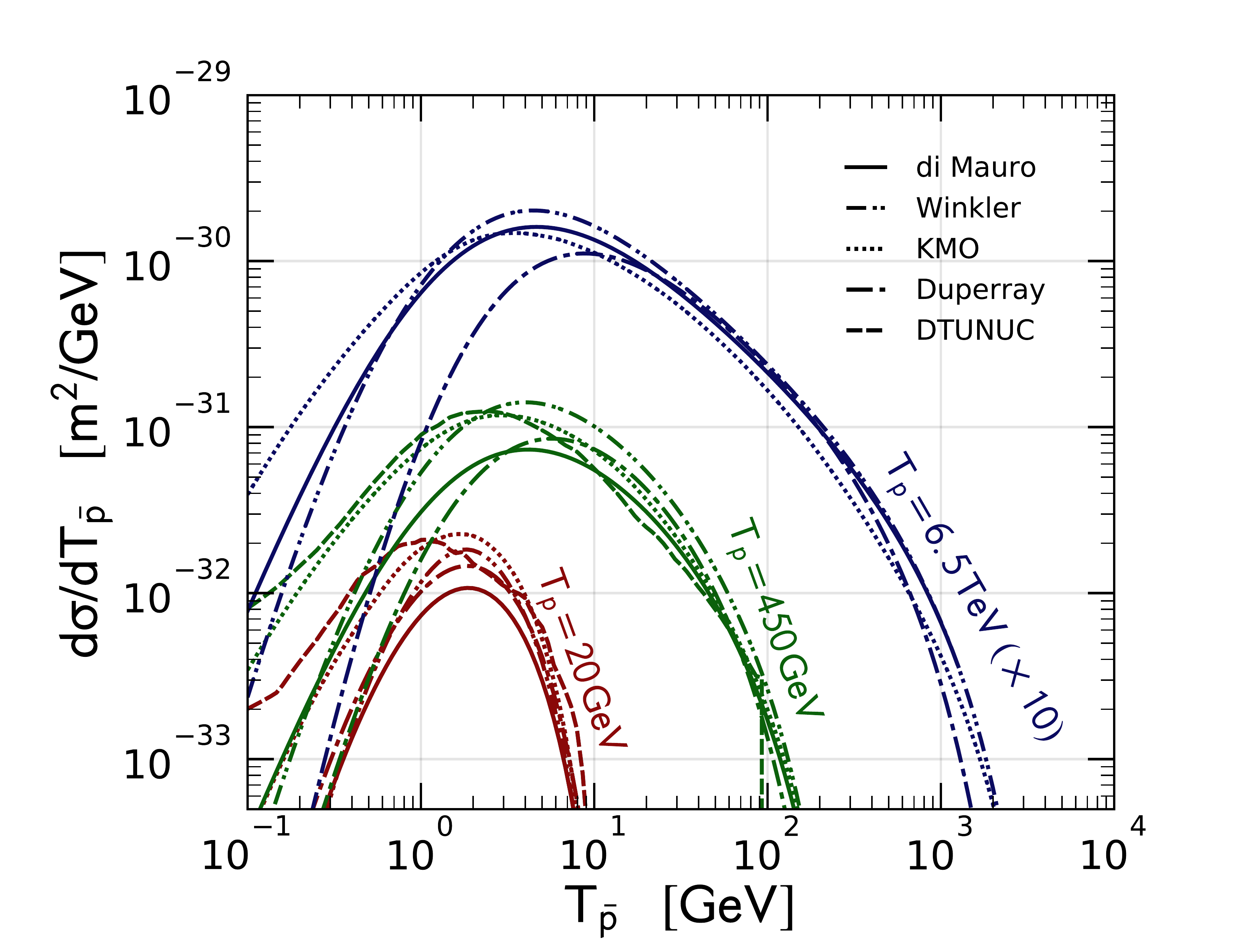}  }
	\caption{Profiles for fixed (a) antiproton and (b) proton kinetic energy 
	         of the $p + p\rightarrow \pb + X$ energy-differential cross 
	         section in the LAB frame from \figref{EnergyDifferentialCS_LAB_diMauro}. 
	         In addition, cross section parametrizations by Tan\&Ng \cite{TanNg:1983_pbarCrossSectionparametrization},
	         Duperray {\textit et al.} \cite{Duperray:2003_pbarCrossSectionparametrizationForPA}
	         (their Eq. 6\noteref{1}),  Kachelriess {\textit et al.} \cite{Kachelriess:2015_pbarCrossSection} 
	         and Winkler \cite{Winkler:2017xor} 
	         are shown for comparison. Panel (c): as panel (b), but for the $p+\mathrm{He}\rightarrow \pb+X$ scattering. 
	         Here we add the DTUNUC parametrization \cite{Simon_Antiproton_CS_Scaling_1998, Donato:2001_DTUNUC}.  }	
	         \label{fig::EnergyDifferentialCS_LAB_profiles}
\end{figure}
\footnotetext{\label{note::1} As discussed in \cite{diMauro:2014_pbarCrossSectionparametrization}, 
                            the parameters $D_1$ and $D_2$ must be interchanged.}

The $\pb$ production cross section is not directly available in the energy-differential form from \eqnref{energyDifferentialCrossSection}, which also enters in \eqnref{sourceTerm_1}. Experiments, rather, measure the angular distribution on top of the energy-differential cross section and then present the Lorentz invariant (LI) form
\begin{eqnarray}
\label{eqn::LICrossSection}
  \sigmaInv (\sS, \xR, \pT)&\equiv&E \frac{d^3\sigma}{d p^3}(\sS, \xR, \pT), 
\end{eqnarray}
where $E$ and $p$ are the total $\pb$ energy and momentum, respectively, $\sS$ is the center-of-mass (CM) energy of the colliding nucleons, 
$\xR = E_\pb^*/E^*_{\pb, {\rm max}}$ (* refers to CM quantities) is the ratio of the $\pb$ energy to the maximally possible energy in the CM frame, 
and $\pT$ is the transverse momentum of the produced antiproton. Note that the three kinematic variables are LI quantities. 
We skipped the subscripts $i,j$ for projectile and target to avoid unnecessarily complicated notation. 
Anyway, \eqnref{LICrossSection} and also the following equations are valid for all combinations of projectile and target, if all quantities are understood in the nucleon-nucleon system.

To relate the LI cross section to the energy-differential one in  \eqnref{energyDifferentialCrossSection}, two steps must be performed. First, the LI kinetic variables $\{\sS, \xR, \pT\}$ need to be related to an equivalent set in the LAB frame, where the target is at rest. 
Typically, the  set is given by the projectile and the $\pb$ kinetic energies, and the scattering angle  $\{T, \Tpbar, \cosT\}$. We give explicit relations in \appref{appendix}.
In a second step, the angular integration has to be performed

\begin{eqnarray}
\label{eqn::energyDifferentialToInvCS}
\frac{d\sigma}{d \Tpbar}(T , \Tpbar) 
  &=&  2\pi   p_\pb  \int\limits_{-1}^{1}           d\cos(\theta) \,\,                             \sigma_\mathrm{inv} \nonumber \\
  &=&  2\pi   p_\pb  \int\limits_{-\infty}^{\infty} d\eta         \,\, \frac{1}{\cosh^2(\eta)} \,  \sigma_\mathrm{inv}.
\end{eqnarray}
Here $\theta$ is the angle between the incident projectile and the produced antiproton in the LAB frame. In the second line of 
\eqnref{energyDifferentialToInvCS} we transform the angular integration to an integration with respect to the 
pseudorapidity defined as 
\begin{eqnarray}
\label{eqn::pseudoRapidity}
  \eta &=& -\ln\left(\tan\left(\frac{\theta}{2}\right)\right).
\end{eqnarray}
This transformation is advantageous because the invariant cross section is very peaked in forward direction at small angles. 
Again, a more detailed derivation of \eqnref{energyDifferentialToInvCS} is stated \eqnref{energyDifferentialToInvCS_full} in the Appendix.
Concerning the limits of the angular integration, we notice that from $\xR\leq 1$ it is possible 
to derive a precise $\theta_{\max}$  or, equivalently, $\eta_{\min}$. 
Nevertheless, in practice it is sufficient to start both integrals 
in \eqnref{energyDifferentialToInvCS} from 0, which is trivially a lower limit.

\begin{table*}[t]
  \caption{Summary of available cross section parametrizations for the antiproton production.}
  \label{tab::CrossSectionparametrization}
  \centering
  \begin{tabular}{l l l l l}
  \hline \hline
  \textbf{parametrization} & \textbf{Species}  & \textbf{Info} &  \textbf{Year} & \textbf{Ref.} \\ \hline   \hline
  Winkler               & $pp$, $p$He,He$p$,HeHe  & analytic & 2017 & 
                  \cite{Kappl:2014_pbarCrossSection,Winkler:2017xor} \\
  di Mauro  {\textit et al.}      & $pp$                    & analytic & 2014 & \cite{diMauro:2014_pbarCrossSectionparametrization} \\
  Duperray  {\textit et al.}      & $pp$, $pA$, $Ap$        & analytic & 2003 & \cite{Duperray:2003_pbarCrossSectionparametrizationForPA}\\
  Kachelriess {\textit et al.}    & $pp$, $pA$, $Ap$, $AA$  & high-energy MC, $\nb$ inclusive, LAB frame & 2015 & 
                  \cite{Kachelriess:2015_pbarCrossSection}\\
  DTUNUC                & $p$He, He$p$, HeHe      & low-energy MC, LAB frame & 1998 & 
                  \cite{Simon_Antiproton_CS_Scaling_1998,Donato:2001_DTUNUC}\\
  Tan\&Ng               & $pp$                    & analytic & 1983 & \cite{TanNg:1983_pbarCrossSectionparametrization}\\
  \hline \hline
  \end{tabular}
\end{table*}

With this information at hand, we  compare three different parametrizations of the $p+p \rightarrow \pb + X$ cross section, as given by 
 \cite{diMauro:2014_pbarCrossSectionparametrization,TanNg:1983_pbarCrossSectionparametrization,Duperray:2003_pbarCrossSectionparametrizationForPA,Winkler:2017xor} and the MC approach in \cite{Kachelriess:2015_pbarCrossSection}
 (see also the information in \tabref{CrossSectionparametrization}). 
\figref{EnergyDifferentialCS_LAB_profiles}  displays profiles of the energy-differential cross section for either fixed $\Tp$ or $\Tpbar$.  At antiproton energies of a few 10 GeV, which are dominantly produced by protons with an energy of a couple of 100 GeV,  all approaches agree well. However, at lower and higher energies  the picture is different. In particular, for antiproton energies below 10 GeV, the deviation between the different approaches is significant. The MC approach in \cite{Kachelriess:2015_pbarCrossSection}, which is intrinsically designed for and trained with high-energy data, is expected to break down below 10 GeV, as clearly visible in the plots. 
Furthermore, the different analytical forms have discrepancies of up to a factor 2 at 1 GeV. We notice that the parametrizations in 
\cite{diMauro:2014_pbarCrossSectionparametrization, Winkler:2017xor} are driven by the NA49 data \cite{NA49_Anticic:2010_ppCollision}, taken at $\sqrt{s}$=17.3 GeV
and covering antiproton energies  from about 8 GeV up to 70 GeV.
The plotted cross sections also include the antiprotons from antineutron decay.  Given the CR propagation timescale,  $\nb$ immediately decays into $\pb$ and has to be included in the antiproton source term. The simplest correction is to multiply the cross section by two. 
However, NA49 data suggest that the $p+p \rightarrow \nb +X$ cross section is larger 
than the $\pb$ one by roughly 50\% around $x_F=0$ \cite{NA49_Anticic:2010_ppCollision}, 
where Feynman scaling variable is defined by $x_F=\pL^*/(2\sqrt{s})$ with the the 
longitudinal momentum $\pL^*$ in the CM frame. 
Hence, in order to properly include the $\nb$ inclusive cross section, 
we multiply the $\pb$ one by 2.3 as suggested by Ref.~\cite{diMauro:2014_pbarCrossSectionparametrization}, 
here and in the following. Ref. \cite{Kachelriess:2015_pbarCrossSection}  
directly provides the $\nb$ inclusive cross section based on their modified 
MC simulations with QGSJET-II and \cite{Winkler:2017xor} gives a specific formula for the isospin violation term.  
Moreover, the parametrization from \cite{Winkler:2017xor} contains explicitly the contribution from hyperons, as it is likely the case also for the MC predictions in 
\cite{Kachelriess:2015_pbarCrossSection}.
We remind here that both the antineutron and 
the hyperon induced antiproton production are still debated and involve large uncertainties. The argumentation relies mostly on
symmetries and theory assumptions, as there are no or few direct measurements, respectively.
Our final results should be understood as requirements not only on the antiproton, but 
also on the antineutron and hyperon production. 
More details might be found e.g. in \cite{diMauro:2014_pbarCrossSectionparametrization, Kappl:2014_pbarCrossSection,Winkler:2017xor}.
In panel (c), we also plot literature results for the $p+\mathrm{He}\rightarrow \pb+X$ channel. 
Concretely, only the MC approach in \cite{Kachelriess:2015_pbarCrossSection} provides both the He channels individually, while the 
parametrization in \cite{Duperray:2003_pbarCrossSectionparametrizationForPA} is fitted to $pA$ data ($p$ as the projectile, nuclei heavier than helium as targets) 
while the one in  \cite{diMauro:2014_pbarCrossSectionparametrization} is only valid for $pp$.
In order to get a He$p$ parametrization form Ref. \cite{Duperray:2003_pbarCrossSectionparametrizationForPA}, 
and for both He channels from the results in  Ref. \cite{diMauro:2014_pbarCrossSectionparametrization}, the following is done.
Naturally, the $p$He cross section  might be Lorentz transformed to He$p$ one. 
However, Ref. \cite{Duperray:2003_pbarCrossSectionparametrizationForPA} already assumes a cross section parametrization with z-symmetry\footnote{Symmetry with respect to the plane perpendicular to the incident particles, namely, no dependence on the longitudinal momentum $\pL$.} in the nucleon-nucleon CM system. 
In this way, we directly conclude that both He channels ($p$He and He$p$) are equal as functions of energy per nucleon. 
For  the up-to-date $pp$ parametrization in \cite{diMauro:2014_pbarCrossSectionparametrization}, 
we extract and translate the $A$ dependence from  \cite{Duperray:2003_pbarCrossSectionparametrizationForPA}. Again \cite{Kappl:2014_pbarCrossSection,Winkler:2017xor} gives its own scaling for the $p$He and He$p$ cross sections.
We see from \figref{EnergyDifferentialCS_LAB_profiles} that different choices for the He cross section lead to significantly different results, in particular below $\Tpbar $ of about 10 GeV, where the discrepancy can reach a factor 10.

\begin{figure}[t!]
	\includegraphics[width=1.\linewidth]
	    {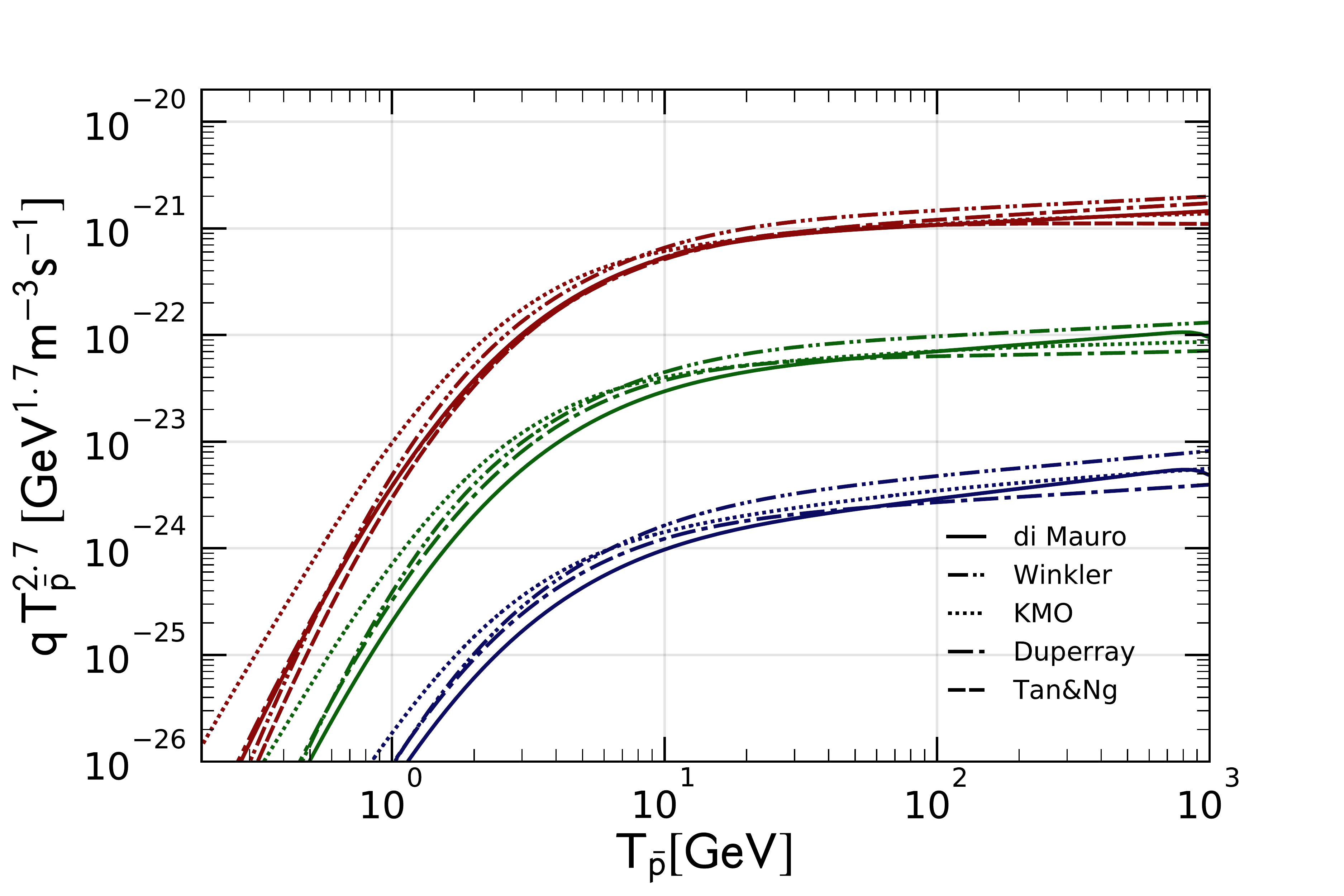}
  \caption{Antiproton source term for CRs of the tree dominant channels $pp$, $p$He, and He$p$ 
           for different parametrizations by Di Mauro {\textit et al.} \cite{diMauro:2014_pbarCrossSectionparametrization}
           (their Eq.~12), Tan\&Ng \cite{TanNg:1983_pbarCrossSectionparametrization}, 
           Duperray {\textit et al.} \cite{Duperray:2003_pbarCrossSectionparametrizationForPA}
           ($pp$ from their Eq. 6\noteref{1} 
           and Eq. 4 for heavier channels), Winkler \cite{Winkler:2017xor},
           and Kachelriess {\textit et al.} \cite{Kachelriess:2015_pbarCrossSection}.
           }
	\label{fig::SourceTerm_differentparametrizations}
\end{figure}

In \figref{SourceTerm_differentparametrizations} we present the source term for 
the three most important production channels $pp$, $p$He, and He$p$. 
Although some of the differences among the cross sections might be integrated out at the source term level (see \eqnref{sourceTerm_1}), 
\figref{SourceTerm_differentparametrizations} shows that, in fact, most of 
the deviations above $\Tpbar=100$ GeV and below 10 GeV remain. 
The predictions from Winkler's parametrization are systematically higher than the other ones above 10 GeV. One possible explanation could be the explicit inclusion of antiprotons 
from the decay of hyperons.
The approach in \cite{Kachelriess:2015_pbarCrossSection} predicts a much larger source term at low energies. 
We are aware of the fact that the heavier elements, namely C and O, contribute to the source term at the percent level. However, the purpose of this analysis is to determine the cross section parameter space of the dominant production channels. In this regard, it is sufficient to consider only $p$ and He each as projectile and target.

The antiprotons, once produced at the source, propagate in the Galaxy experiencing magnetic diffusion together with minor effects 
that shape the flux measured at the Earth. Since the source spectrum is the fundamental quantity with which we will work, we do not 
pause here on the propagation details. They can be found, {\it i.e.}, in \cite{Donato:2001_DTUNUC}, where the CR propagation is considered within a semianalytical diffusive model.  However, errors induced by propagation modelings  can be considered reasonably negligible when dealing with relative flux uncertainties, as we will do in the following of our paper. 
In the last years, more complex scenarios have been proposed with the main aim of explaining the rising positron fraction observed by PAMELA and AMS-02.  
In \cite{Blasi:2009bd,Mertsch:2014poa}, it is predicted that secondary CRs receive a primary contribution from scatterings of primary nuclei inside the supernova remnants, before being propagated. 
Refs.  \cite{Tomassetti:2012ga,Tomassetti:2015mha} speculate that 
diffusion might be different close to the Galactic disk and further away, and build  a  two-halo diffusion model. 
In \cite{Cowsik:2016wso}, the secondary origin of antiprotons in the nested leaky-box model is discussed. 
However, as discussed in the next section, our choice to work with relative (and not absolute) flux uncertainties strongly limits 
the effects of alternative production and propagation modeling on our results on the production cross sections.

\section{\label{sec::methods} Methods for determining the precision on the cross section}

The general idea of the present analysis is to determine the uncertainty requirements on $\pb$ cross section measurements for a given cosmic antiproton flux accuracy. 
Firstly, we determine the contribution to the source term from each point in the parameter space of the fully differential LI cross section.
Then we derive the uncertainty requirements on cross section measurements according to two principles:
(i) the total uncertainty shall match experimental flux accuracy dictated by AMS-02, which provides the currently most precise measurement, and
(ii) in the parameter space regions, where the cross section provides a dominant contribution to the source term, 
we require higher accuracy. 
In the following we will provide a detailed explanation of our strategy.

We start from the uncertainty level in the antiproton flux measurement. 
Our prior is the AMS-02 data, which display the most accurate determination over the widest energy range (see \figref{FluxMeasurements_AMS-02}). 
This TOA spectrum of AMS-02 has to be related to the $\pb$ source spectrum.
First of all, we have to extract the local IS flux from the 
TOA one. We choose to correct the effect of the solar modulation by means of the force-field approximation, 
dictating to simply shift all data points by $\Delta E = |Z| \phi_\odot$, having fixed  $\phi_\odot= 600$ MV.
Then, the relation between the IS flux and the source spectrum is given by propagation in the Galaxy, 
which is usually described by a diffusion equation (see \cite{Donato:2001_DTUNUC} and refs. therein).

As a first approximation, it is reasonable to assume that relative uncertainties of IS flux and the source term are equal above 
$\sim1$~GeV. Flux and the source term are linked by linear differential equation. Thus, the main term of this 
equation, namely, the diffusion term, indeed keeps the ratios unaffected. 
The only possibility are energy distortions between the propagated flux and the source spectrum which may arise from 
reacceleration at very low energies.
However, we explicitly checked by propagating several, strongly peaked, toy-source-term spectra 
and comparing them with the resulting propagated flux
that the energy distortion is negligible down to 1~GeV.
The baseline for the cross check is the \textsc{Galprop}-based global analysis of CR propagation performed in 
\cite{2016PhRvD..94l3019K}  (and refs. therein).
Conclusively, we use the relative flux uncertainty $\sigma_{\phi_\pb}$ as proxy of the source term uncertainties ${\sigma_q}$:
\begin{eqnarray}
  \sigmaRel_q (\Tpbar) \equiv \frac{\sigma_q(\Tpbar)}{q(\Tpbar)}  \approx \frac{\sigma_{\phi_\pb}(\Tpbar)}{\phi_\pb(\Tpbar)}.
\end{eqnarray}
\begin{figure}[t!]
	\includegraphics[width=1.\linewidth]
	    {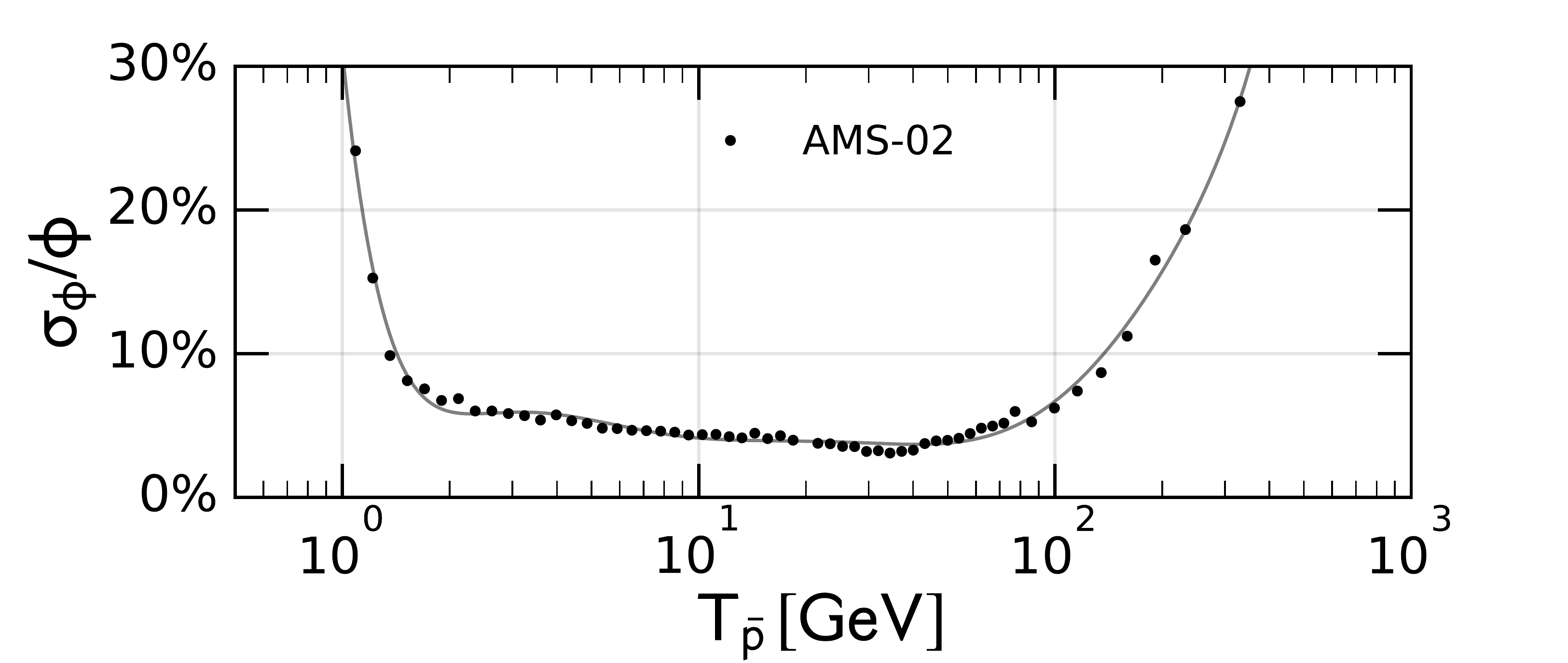}
  \caption{Relative uncertainty of the AMS-02 antiproton flux measurement \cite{AMS-02_Aguilar:2016_AntiprotonFlux}.
           We show the demodulated IS flux, in other words, all data points are shifted by 600 MeV. }
	\label{fig::AMS-02_pbar_uncertainty}
\end{figure}
The quantity $\sigmaRel_q (\Tpbar)$ can be read from \figref{AMS-02_pbar_uncertainty}, where it has been derived from the 
 AMS-02 $\pb$ measurements \cite{AMS-02_Aguilar:2016_AntiprotonFlux}. 
 From this figure we can clearly see the precision level of the current AMS-02 data. In particular, it is about 5\%  between 1 and 100 GeV. This is the minimum level of accuracy which is required to any prediction. The uncertainty of the source term has to be distributed on the single production channels. 
 We will assume that the relative uncertainties in all the $p$ and He channels are equal, meaning
\begin{eqnarray}
  \sigmaRel_{q}  (\Tpbar) = \sigmaRel_{q_{ij}}  (\Tpbar).
  \label{eqn::unc_channels}
\end{eqnarray}


To determine the contribution from each parameter point of the invariant cross section 
we work with the full expression of the source term. 
Inserting \eqnref{energyDifferentialToInvCS} in \eqnref{sourceTerm_1} and changing the energy 
integration to $\log(T)$ results in
\begin{eqnarray}
\label{eqn::sourceTerm_2}
  q(\Tpbar) &=&  \underset{\log(E_{\rm th})\qquad}{\int\limits^\infty d\log(T)} \int\limits_{0}^{\infty} d\eta   
          \underbrace{ \frac{8\pi^2 \, p_\pb \,n_{\mathrm{ISM}}\,T\, \phi(T) \, \sigmaInv(\Tpbar,T,\eta)}{\cosh^2(\eta)} }_{\equiv I(\Tpbar, T, \eta)}, \nonumber \\
         & \, &
\end{eqnarray}
where we have dropped the labels $i,j$ indicating the species of the incoming CR flux and ISM. 
We then define the containment
\begin{eqnarray}
\label{eqn::containence}
  x(\Tpbar, T, \eta) &=&  \frac{1}{q(\Tpbar)} \underset{I(\Tpbar, T', \eta')>I(\Tpbar, T, \eta) \qquad \qquad}
                                {\int d\log(T')  d\eta'\,\, I(\Tpbar, T', \eta')}. 
\end{eqnarray}
The integrand of the source spectrum $I(\Tpbar, T, \eta)$ has been defined in \eqnref{sourceTerm_2}. 
The containment function varies between $[0,1]$ and has to be understood as follows. 
The parameter space with, e.g., $x(\Tpbar, T, \eta)<0.9$ contains 90\% of the source term at the given antiproton energy.
To be more precise, we calculate the smallest containment areas in the variables $\log(T)$ and $\eta$,  
for each given $\Tpbar$. In practice, this is obtained by calculating the integral from \eqnref{containence} in the following way. 
On the two dimensional grid at fixed $\Tpbar$, we sum the integral of $I(T, \Tpbar, \eta)$ over the  $\log(T)$ and $\eta$ bins starting from the largest to the smallest contribution, until the required value of $x(T, \Tpbar, \eta)$ is reached.
Although there is a freedom to choose the exact set of kinetic parameters, the general behavior of containment contours is not largely affected, because the integrand $I$ is a strongly peaked function. 
Furthermore, cross sections and source terms are power laws. The choice of logarithmic variables, $\log(T)$ and $\eta$, respects their natural scaling properties.

\begin{figure}[t!]
  \includegraphics[width=1.\linewidth]
	    {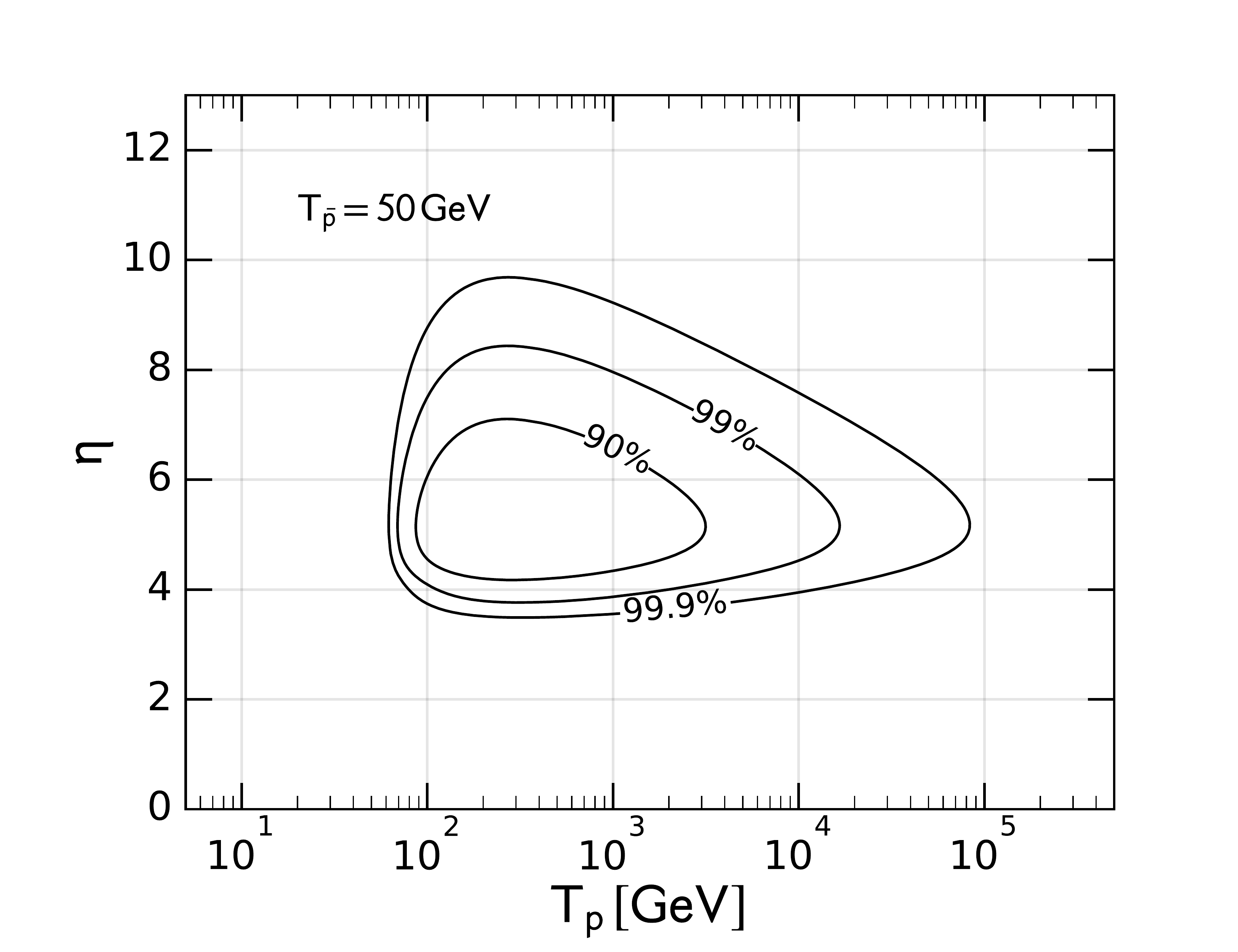}
  \caption{Isocontours for the containment function  $ x(\Tpbar, T, \eta)$ for the 
            contribution of the $pp$ channel to the antiproton source term as function of the kinetic variables in the LAB frame, 
            namely, the kinetic energy of the incident proton $\Tp$ and pseudorapidity of the antiproton $\eta$, 
            at a fixed antiproton kinetic energy $\Tpbar=50$ GeV. A 90\% level corresponds to $x$=0.9.
            }
	\label{fig::Contribution_Tp_eta__LABframe}
\end{figure}

\figref{Contribution_Tp_eta__LABframe} illustrates the containment parameter space for the $pp$ channel 
when $ x(\Tpbar, T, \eta)$ = 0.90, 0.99 and 0.999, as function of $\eta$ and $\Tp$ at fixed $\Tpbar=50$ GeV. 
Here and when not differently stated, we assume the cross section parametrization as in Ref. \cite{diMauro:2014_pbarCrossSectionparametrization}.
However, we do not expect large systematic deviations by changing the underlying cross section parametrization. We will explicitly verify this by changing to the Duperray 8\ et al. \cite{Duperray:2003_pbarCrossSectionparametrizationForPA} or Winkler  \cite{Winkler:2017xor} expression below. 
Note, that it is not possible to check any of the MC results (like for example \cite{Kachelriess:2015_pbarCrossSection})
because they are not available with complete angular dependence.
In \figref{Contribution_Tp_eta__LABframe} we can see that the 90\% of the antiprotons at $\Tpbar$ = 50 GeV 
are produced by protons with energies spanning about 90 GeV - 3 TeV and $\eta$ between 2 and 7, depending on $\Tp$. 
In order to include the 99.9\% of the $pp$ source spectrum, one has to consider protons with energies up to 70 TeV with 
pseudorapidity values around 5.

From the containment $x$ it  is easy to assign values for $\sigmaRel_{\sigmaInv}$. 
Since increasing $x$ corresponds to decreasing contribution by construction, $\sigmaRel_{\sigmaInv}$ shall increase while $x$ varies from 0 to 1. 
By increasing $x$ up to 1, one increases the spanned kinematical parameter space.
This correspondence ensures that our initial requirement (ii), see beginning of \secref{methods}, namely, more accurate determinations of the cross sections in the parameter space with large contribution, is fulfilled.

\begin{table}[b!]
  \caption{Grid properties for the LAB and CM frames variables adopted in our  numerical calculations.}
  \label{tab::GridProperties}
  \centering
  \begin{tabular}{l r l c r l l}
  \hline \hline
  \textbf{Variable [unit]} & \multicolumn{5}{l}{\textbf{Range}}                 & \textbf{Size/Scale} \\ \hline   \hline
    $T$ [GeV/n]            & $5$&$ \times 10^{+0}$ & -- & $5 $&$\times 10^{+5}$  & 400/log\\
    $\Tpbar$ [GeV]         & $1$&$ \times 10^{-1}$ & -- & $1 $&$\times 10^{+3}$  & 400/log \\
    $\eta$                 & $0$&$               $ & -- & $13$&$              $  & 400/linear \\ \hline
    $\sS$ [GeV]            & $1$&$ \times 10^{+5}$ & -- & $1 $&$\times 10^{+5}$  & 400/log \\
    $\pT$ [GeV]            & $1$&$ \times 10^{-2}$ & -- & $1 $&$\times 10^{+2}$  & 400/log \\
    $\xR$                  & $0$&$               $ & -- & $1 $&$              $  & 400/log \\
  \hline \hline
  \end{tabular}
\end{table}
We fix a precision level for the relative uncertainties on the LI cross section for $\pb$ production.  
To keep things simple we choose the step function
\begin{eqnarray}
\label{eqn::step_function}
  \sigmaRel_{\sigmaInv}(x, \Tpbar) = 
    \begin{cases}
      3\%  & x< x_t(\Tpbar) \\
      30\% & \text{elsewhere},
    \end{cases} 
\end{eqnarray}
where $x_t(\Tpbar)$ is a threshold value for the containment function. 
The values of 3\% and 30\% are a free choice, which, however, is suggested by by the most precise values of current cross section measurements (see \figref{DataComparison_NA49} in the following), on the one hand, and by the spread of various parametrizations in the energy regime of interest, on the other hand.
Anyway, we will provide a comparison with different assumptions for the step function in \secref{results}.
Finally, the threshold $x_t(\Tpbar)$ is fixed by the requirement to match AMS-02 accuracy, which is guaranteed by solving  
\begin{eqnarray}
\label{eqn::fix_threshold}
  \int\limits_0^1 dx \,\, \sigmaRel_{\sigmaInv}(x, \Tpbar) = \sigmaRel_q (\Tpbar).
\end{eqnarray}
In practice, the right-hand side of this equation is taken from the parametrization of the uncertainties on the $\pb$ data as reported in
\figref{AMS-02_pbar_uncertainty}. 

This procedure allows us to derive the required levels for $\sigmaRel_{\sigmaInv}$. 
Technically, we determine $\sigmaRel_{\sigmaInv}$ on a $400^3$ grid in the LAB frame variables summarized in \tabref{GridProperties}. 
To get the distribution in $\sS, \xR, \pT$ we transform those three variables to $T, \Tpbar, \eta$ and then interpolate on our grid.
To get a bijective mapping we have to add an assumption on the sign of $\pL$. 
So we get two values for each set of $\sS, \xR, \pT$ and choose the minimum.

\begin{figure*}[t]
	\subfloat[LAB frame]{  \includegraphics[width=0.45\textwidth]{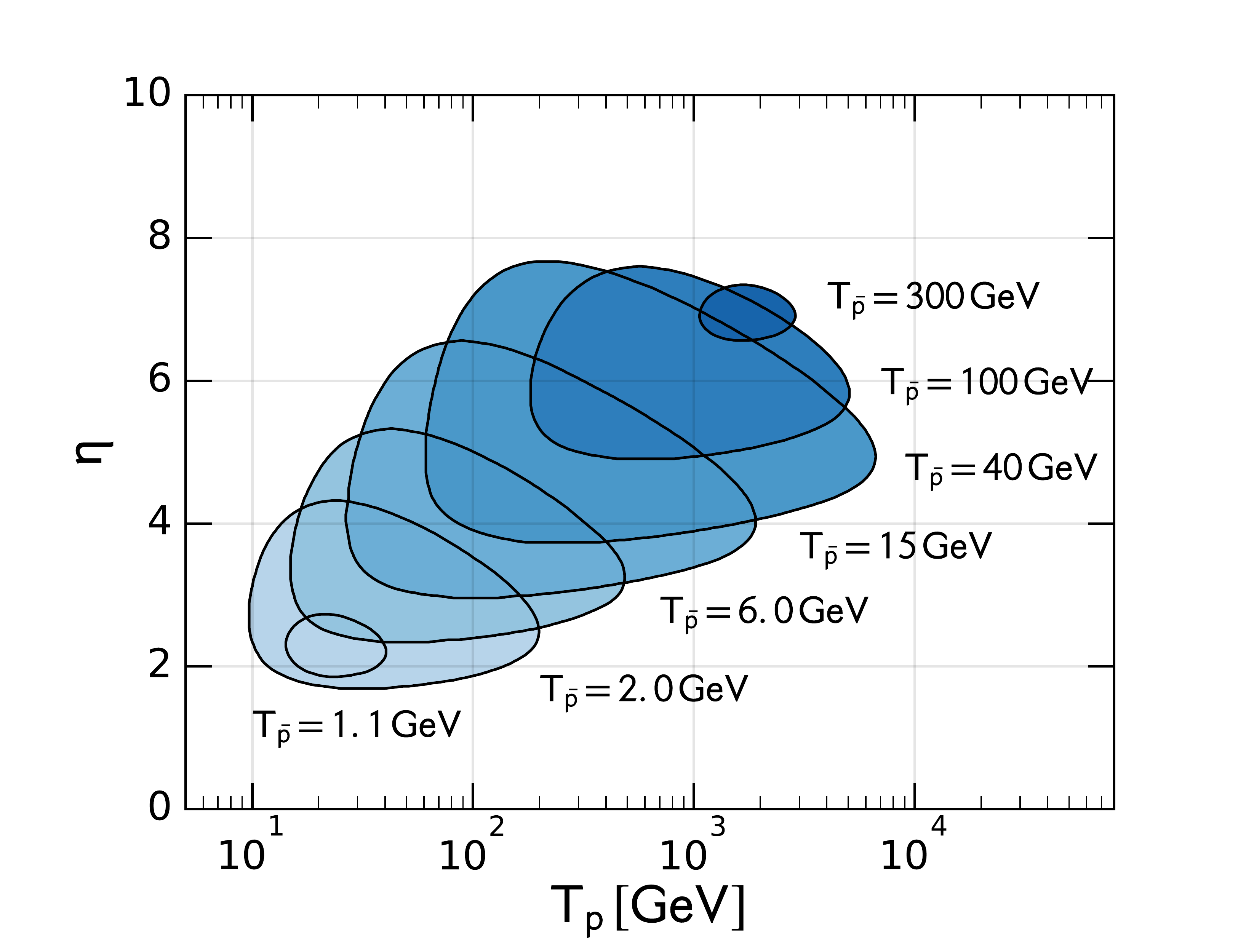}   }
	\hspace{0.05\textwidth}	
	\subfloat[CM frame ]{  \includegraphics[width=0.45\textwidth]{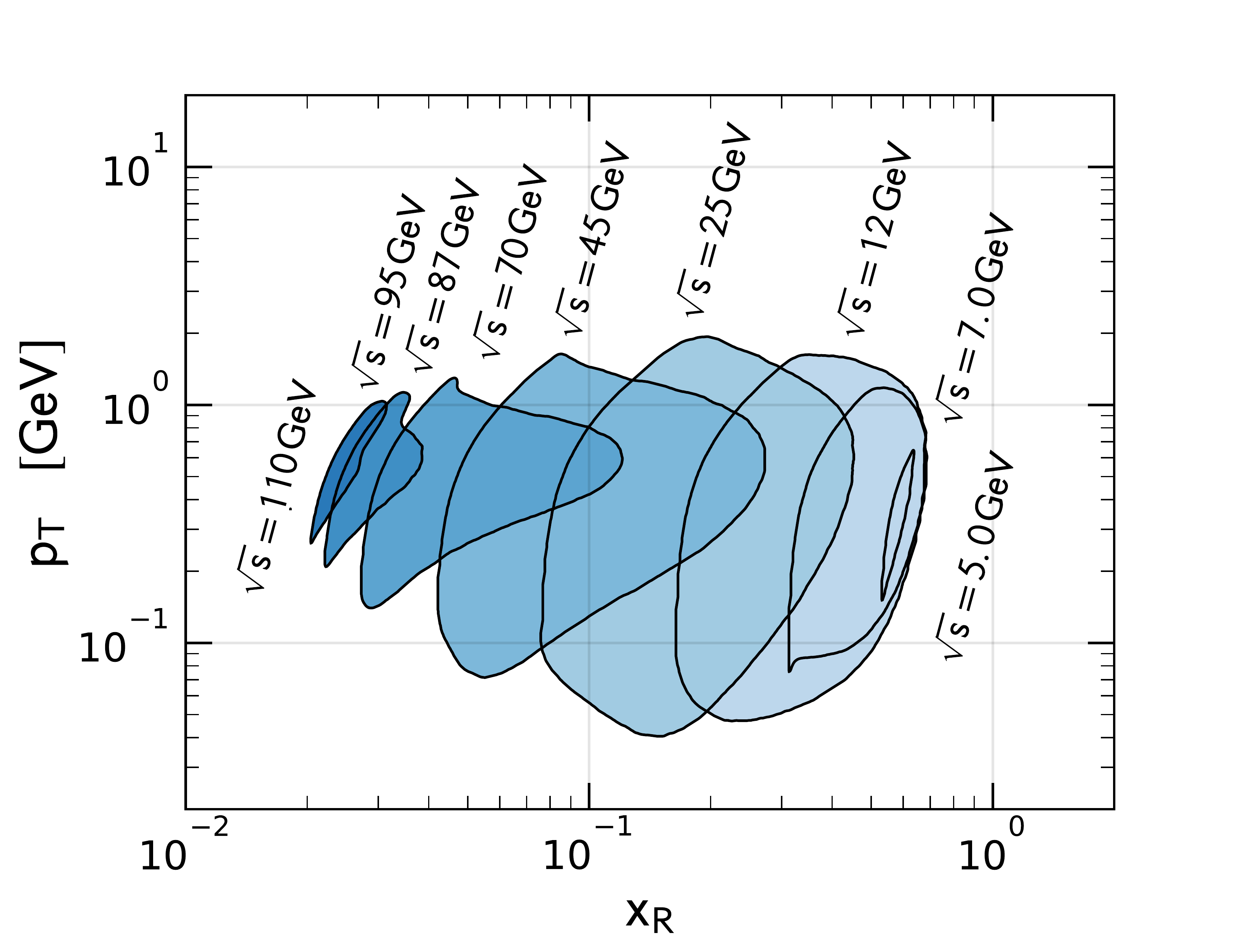}    }
	\caption{Parameter space of the $pp$ to $\bar{p}$ cross section necessary to determine 
	         the antiproton source term with the accuracy reached by recent AMS-02 measurements 
	         \cite{AMS-02_Aguilar:2016_AntiprotonFlux}. 
	         Here we require that  the cross section has to be known by 3\% within the blue shaded regions
	         and by 30\% outside the contours. The left (right) panel displays the result for the LAB (CM) 
	         reference frame variables.  }
	\label{fig::ParameterSpace_AMS-02}
\end{figure*}

\section{\label{sec::results}Results}
We derive the parameter space of the inclusive $\bar{p}$ cross section which should be covered 
 to determine  the antiproton source term with the accuracy dictated by recent AMS-02 measurements 
\cite{AMS-02_Aguilar:2016_AntiprotonFlux}. 
We show our results as functions of the kinematical variables in both the LAB and CM reference frames. 
As explained in the previous section, in the LAB frame the parameter space is described by the kinetic energy of the proton $\Tp$, the kinetic energy of the antiproton $\Tpbar$, and and pseudorapidty of the antiproton $\eta$. 
Equivalently, the results can be expressed in terms of the CM frame, given by the CM energy $\sqrt{s}$, the ratio between the antiproton energy and its  maximal energy $x_R$, and the transverse momentum of the antiproton $\pT$. 
Unless stated differently, the $pp$ cross section parametrization is chosen as in Di Mauro {\textit et al.} \cite{diMauro:2014_pbarCrossSectionparametrization}.

\figref{ParameterSpace_AMS-02} shows the parameter space that has to be covered in order to guarantee the AMS-02 precision level 
on the $\pb$ source term, if the 
$p+p\rightarrow \bar{p}+X$ cross section is determined with 3\% uncertainty within the blue shaded regions and by 30\% outside the contours. 
The plot is done for the LAB (left panel, a) and CM (right panel, b) reference frame variables. For the LAB frame
we show the contours as functions of $\eta$ and $T$, for selected values of $\Tpbar$ from 1.1 (the lowest energy below 30\% uncertainty in the CR $\pb$ flux, see \figref{AMS-02_pbar_uncertainty}) to 300 GeV. As expected the contour size decreases when $\Tpbar$ approaches to 1 GeV, because there the AMS-02 uncertainty on the antiproton flux reaches 30\%.
A similar explanation holds for large $\Tpbar$.
Antiprotons of increasing energy require the coverage of increasing $\eta$ values. 
For example, $\sigma_{\rm inv}(p+p\rightarrow \pb+X)$ at $\Tpbar$=2 GeV is known at 3\% level if data were taken with proton beams 
between 10 and 200 GeV and pseudorapidity from 1.8 to 4. 
If the whole AMS-02 energy range had to be covered with high precision, one should collect $p+p \rightarrow \pb+X$ cross section data with proton beams from 
10 GeV to 6 TeV, and $\eta$ increasing from 2 to nearly 8. 
\figref{ParameterSpace_AMS-02}b displays analogous information in the CM reference frame. We fix $\sqrt{s}$ to  
representative values from 5 to 110 GeV, and identify the regions in the $\pT$-$\xR$ plane.
A full coverage of the parameter space should scan $\pT$ from 0.04 to 2 GeV, and $\xR$ from 0.02 to 0.7. For each value of $\sqrt{s}$, the extension of the contour within which 
the cross section is required at 3\% precision level is correlated to the AMS-02 uncertainty on the antiproton flux. 

\begin{figure*}[t!]
  \subfloat[LAB frame]{  
    \includegraphics[width=.45\textwidth]
	    {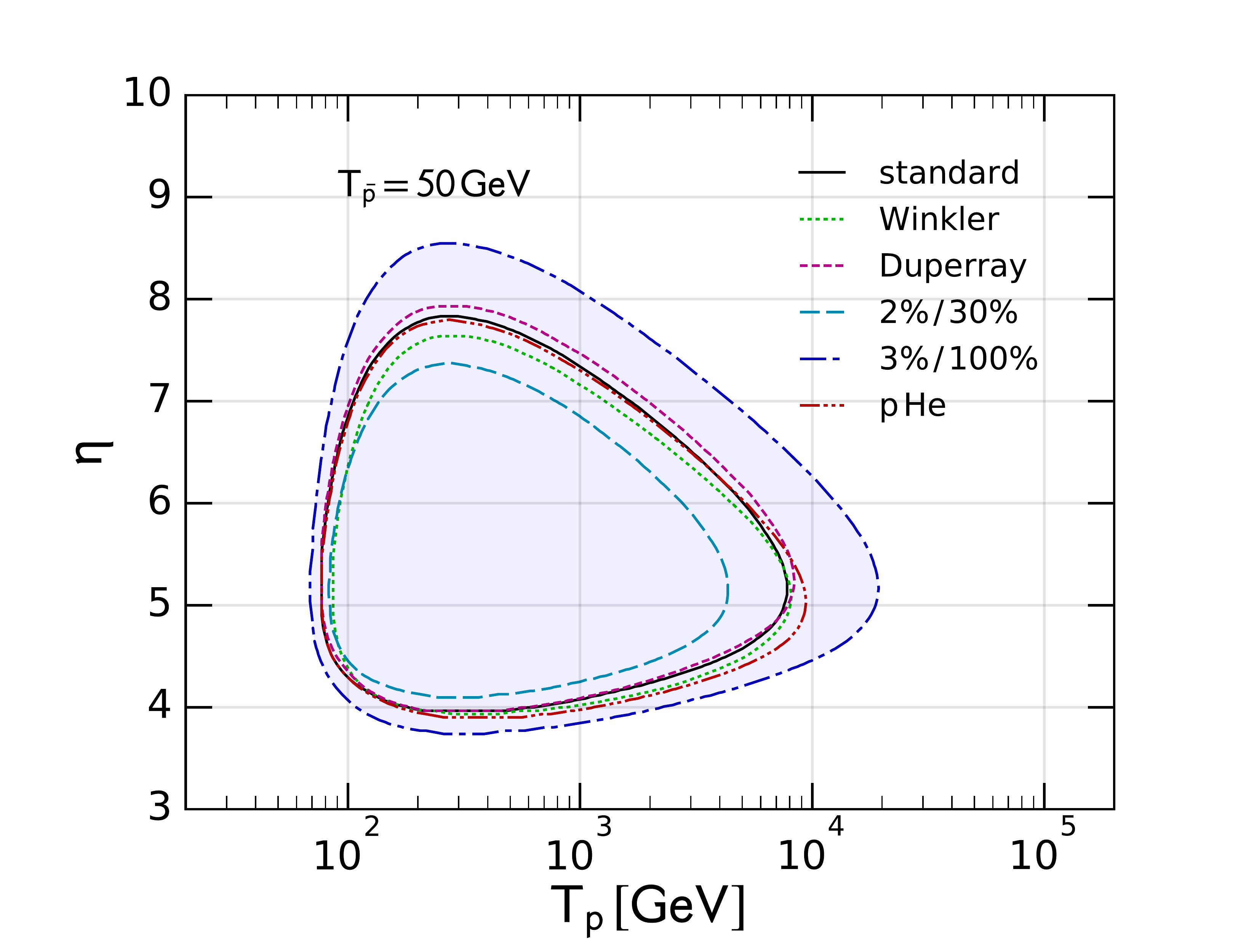}
	}
	\hspace{0.05\textwidth}	
	\subfloat[CM frame]{  
    \includegraphics[width=.45\textwidth]
	    {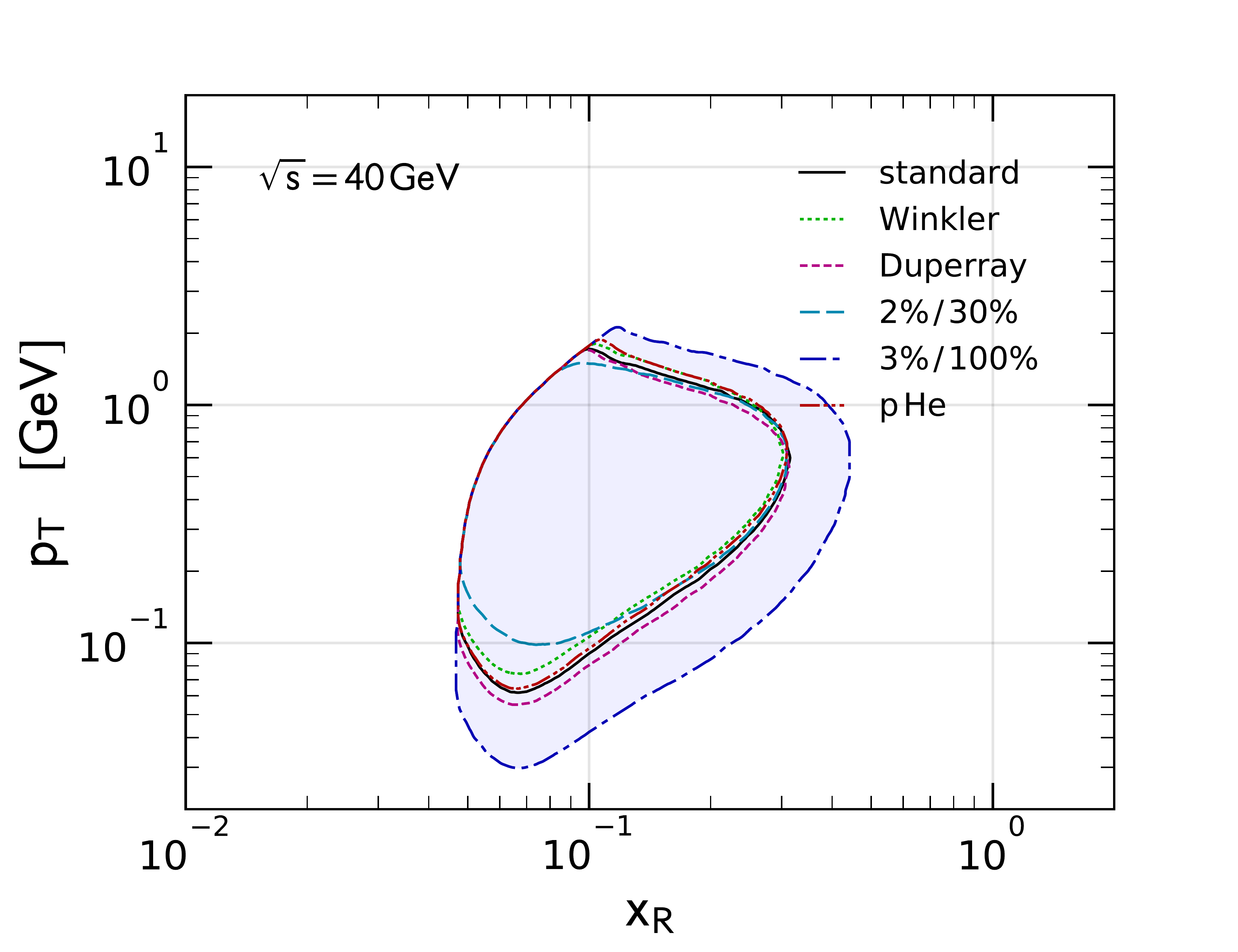}
  }  
  \caption{Different systematic setups are compared (a) in the LAB frame, exemplary  at $\Tpbar=50$ GeV, 
           and (b) the CM frame at fixed $\sS=40$ GeV.
           The standard setup with the $pp$ cross section by Di Mauro 
           \cite{diMauro:2014_pbarCrossSectionparametrization} and inner/outer uncertainty 3\%/30\%, 
           is altered in one of the following details: 
           (i)   the $pp$ cross section is replaced by the Winkler           
           \cite{Winkler:2017xor} (dotted line) or Duperray parametrization
           \cite{Duperray:2003_pbarCrossSectionparametrizationForPA} (their Eq. 6) (dashed line), 
           (ii)  inner/outer uncertainty are replaced by 2\%/30\% (long-dashed line) or 3\%/100\% (dot-dashed line), and
           (iii) instead of $pp$ the $p$He channel is considered (double-dot-dashed line).}
	\label{fig::Comparison_setups}
\end{figure*}

Furthermore, we estimate possible systematic effects on our predictions. In \figref{Comparison_setups} we present the same information as in \figref{ParameterSpace_AMS-02}, now
with different parametrizations for the cross section and modified requirements on the uncertainty levels. We recall that our standard setup is fixed by the $\sigma_{\rm inv}(p+p\rightarrow \pb+X)$ 
as in Ref. \cite{diMauro:2014_pbarCrossSectionparametrization} and uncertainty requirements of 3\%/30\% (inner/outer regions) as 
in \eqnref{step_function}. In \figref{Comparison_setups}a we display the results in the LAB reference frame for $\Tpbar$=50 GeV. 
Changing the parametrization for the cross section to the ones in \cite{Winkler:2017xor} or \cite{Duperray:2003_pbarCrossSectionparametrizationForPA} (their Eq. 6) has negligible effects.
Instead, a smaller (higher) inner $ \sigmaRel_{\sigmaInv}$ implies a smaller (larger)  contours. 
Moving from 2\%-30\% to 3\%-100\%, the $\Tp$ needed coverage pushes 4~TeV to 20~TeV.
As a final point in this figure we address the $p$He cross section, where the computation is performed in the standard setup. It is interesting to note that the covered parameter space is very similar to $pp$. This result is expectable because at first order the $p$He and $pp$  cross sections are simply related by a rescaling, which drops out of the calculation. 
In this regard, it is possible to interpret all plots with $pp$ results also for the heavier channels $p$He or He$p$.
Note, however, that in this case $\Tp$ has to be understood as the kinetic energy per nucleon of the projectile. 
Naturally, all considerations also hold for the representation in the CM reference frame, as shown in 
\figref{Comparison_setups}b.

\begin{figure}[t!]
  \includegraphics[width=1.\linewidth]
	    {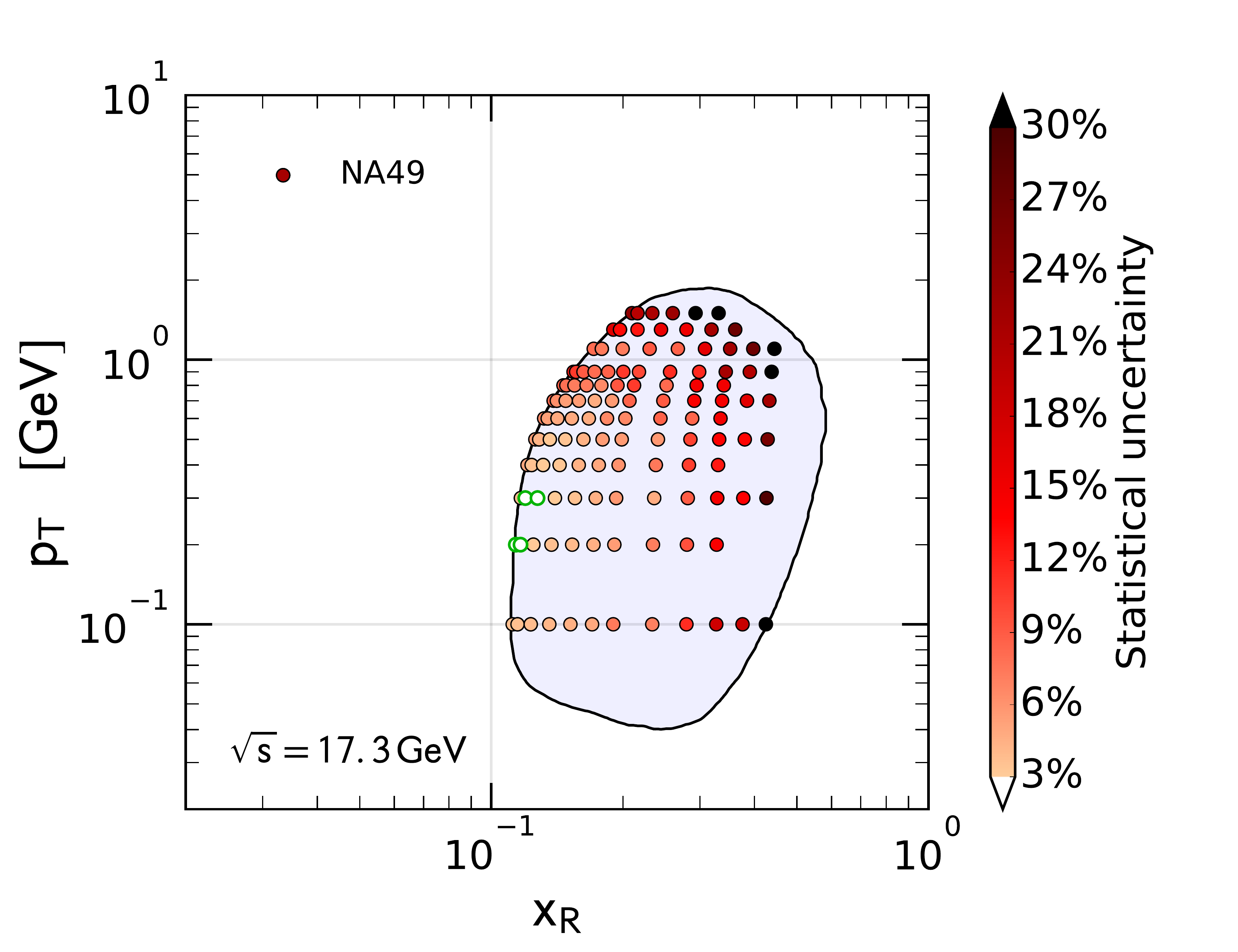}
  \caption{Comparison of the statistical measurement uncertainty by the experiment NA49
           \cite{NA49_Anticic:2010_ppCollision}
           to our standard requirement, 3\% accuracy inside the contour and 30\% outside, which is shown by color grades.
           The points with green border fulfill our 3\% requirement. }
	\label{fig::DataComparison_NA49}
\end{figure}
In \figref{DataComparison_NA49} we display the state of the art by showing the NA49 data \cite{NA49_Anticic:2010_ppCollision} 
on the $\sigma_{\rm inv}(p+p\rightarrow \pb + X)$. Data is taken at $\sqrt{s}=17.3$ GeV corresponding to a proton beam at 158 GeV momentum on a fixed proton target. 
The points show the statistical uncertainty in the CM parameter space, while an overall systematic error of 3.3\% (quadratic sum) is omitted. 
 The parameter space of NA49 matches the requirement of AMS-02 coverage. But only four data points, indicated by the green border, fulfill the uncertainty requirement. The other data points exceed the 3\% error level even by a large amount. 
We notice that due to the high density of points, which can improve the shape determination of $\sigmaInv(p+p\rightarrow\pb+X)$, the statistical uncertainties might decrease after the integration leading to the source term. 
However, it is not straightforward to estimate the amount of this improvement, if any, as the cross section has to be folded with the primary fluxes of $p$ or He, which are grossly
power laws with spectral index 2.7. 
On the other hand, the 3.3\% overall systematic uncertainty is irreducible.   
Similar information for a selection of data samples collected in the previous decades are shown in \appref{old_data}.

\begin{figure}[b!]
  \includegraphics[width=1.\linewidth]
	    {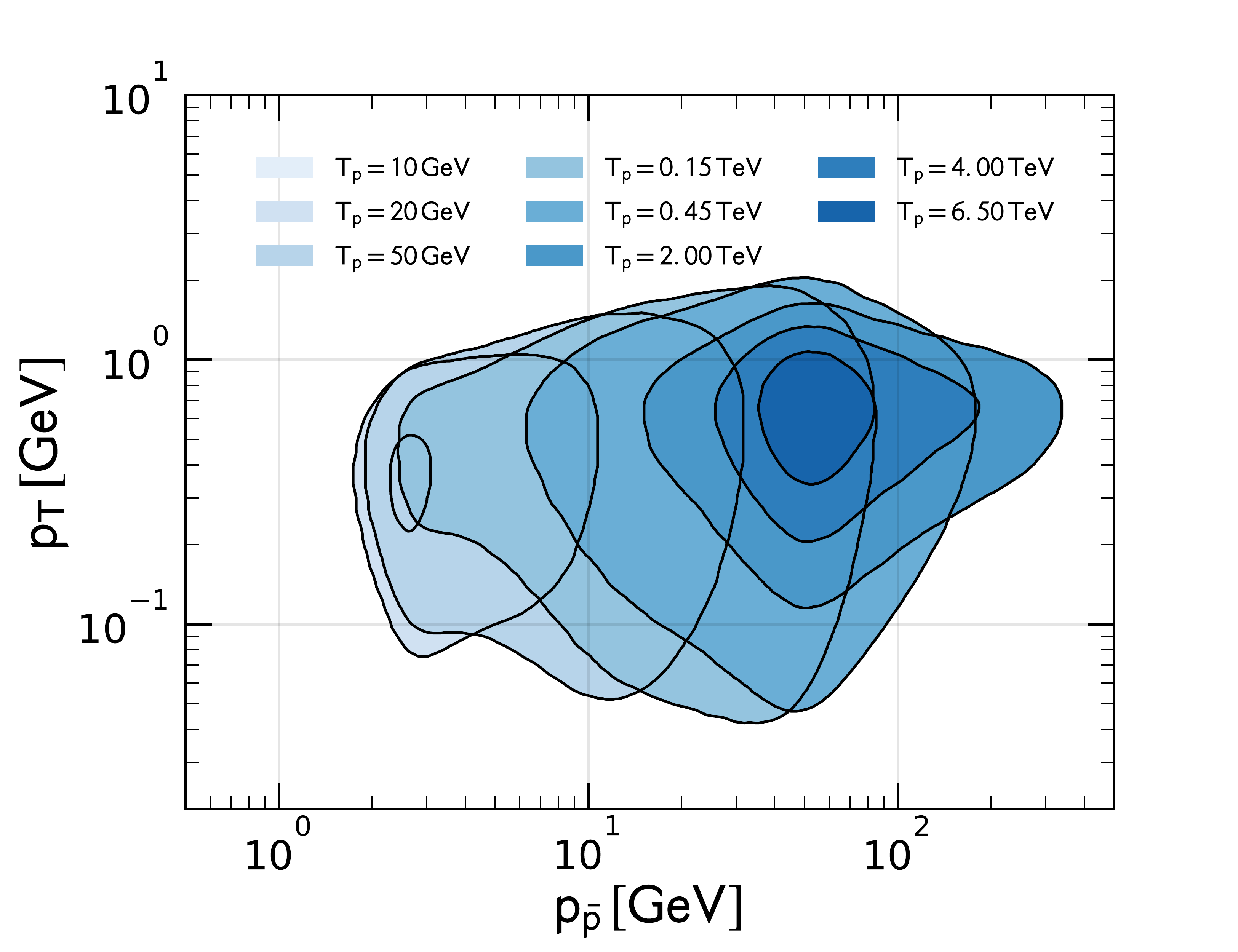}
  \caption{Parameter space for the $p$He channel 
            corresponding to an exemplary fixed target experiment. 
            The different shaded areas correspond to different proton 
            beam energies. }
	\label{fig::ParameterSpace_AMS-02__LHCb_profiles}
\end{figure}

Given the relevance of the helium production channels in the $\pb$ spectrum, we come back to the discussion of heavier channels and explicitly study the parameter space  an experiment, using high-energy protons 
scattering off a fixed helium target, should have in order to fulfill the 3\%-30\% requirement and reach the AMS-02 precision of the $\pb$ spectrum. 
We remind that, as stated in \eqnref{unc_channels}, the relevant cross section uncertainties are set equal for all the production channels.  
In \figref{ParameterSpace_AMS-02__LHCb_profiles} we display the according contours in the parameter space of antiproton momentum and transverse momentum. All variables are in the fixed target frame and, hence, the conveyed information is very similar to \figref{ParameterSpace_AMS-02}(a).
The parameter space which has to be measured spans from below 10 GeV to more than 6.5 TeV, while the required $\pb$ momentum tracks  the AMS-02 measurement range from about 1 to 350 GeV. The LI transverse momentum $\pT$ remains, as expected from \figref{ParameterSpace_AMS-02}(b), between 0.04 and 2 GeV. Again, at $\Tp$ below 10 GeV or equivalently above 2 TeV the size of the contours shrinks because then the dominant production of antiproton is below or above the AMS-02 measurement range, respectively. 

\begin{figure}[t!]
  \includegraphics[width=1.\linewidth]
	    {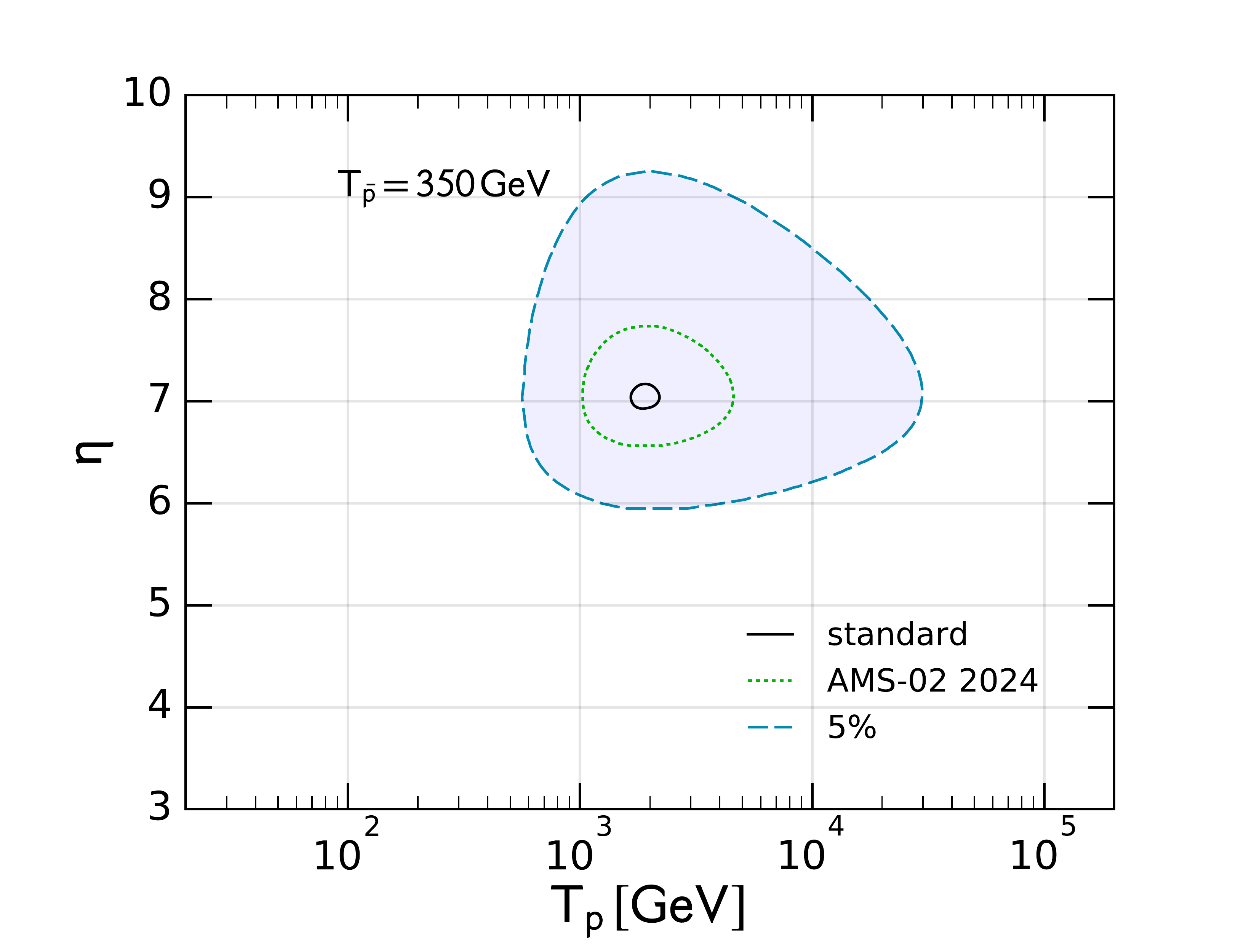}
  \caption{Future projection. AMS-02 in 2024 and 5\% constant uncertainty.}
	\label{fig::Comparison_HE}
\end{figure}
Finally, we undergo the  exercise to investigate future requirements on cross section measurements to explain the highest energies which is measured by AMS-02. 
\figref{Comparison_HE} is a dedicated plot for fixed $\Tpbar=350$ GeV, which corresponds to the central value of the last energy bin 
 of the current AMS-02 measurement. The figure contains the same information as \figref{ParameterSpace_AMS-02}(a), displaying the usual contours with 3\% and 30\% accuracy requirement for inner and outer parameter space, respectively.
We find that the $p+p\rightarrow \pb+X$ cross section should be measured with a proton beam of $\Tp\sim2$ TeV and $\eta$ around 7. 
The corresponding CM frame variables are $\sqrt{s}\sim 60$ GeV, $\xR \sim 0.18$, and  $\pT \sim0.64$ GeV, which are probably 
more feasible for experiments than the high pseudorapidity values  required in the LAB frame. 
\\
In the present situation the large energy  part $\Tpbar\gsim 100$ GeV of the AMS-02 flux is dominated by statistical uncertainties.
Hence, until the scheduled operation of AMS-02 in 2024, the accuracy will increase. To estimate this improvement, we assume that systematical uncertainties stay constant, while the statistical uncertainty reduces according to the Poisson statistics. The current AMS-02 analysis contains four years of data, and until 2024 AMS-02 may collect 13 years of data. Accordingly, we rescale the statistical error by a factor of $\sqrt{4/13}$. As expected, the contour in \figref{Comparison_HE} slightly increases. As a last step, we show the size of a contour if we require 5\% accuracy on the whole AMS-02 energy range. This leads to a significant increase of the 
$\Tp-\eta$ contour. Now fixed target measurements would need to cover a parameter space for beam energies from about 0.5 TeV up to  30 TeV,  and for pseudorapidity from 6 to above 9.

\section{\label{sec::conclusion}Summary and Conclusion}
Antiprotons in CRs have been measured with space-based experiments with unprecedented accuracy. 
After the high-precision measurement by the PAMELA detector, the flux of cosmic-ray (CR) antiprotons has been provided with unprecedented accuracy by AMS-02 on the International Space Station, with data in the energy range 0.5-400 GeV and errors 
that are as low as 5\%.
The CR antiprotons are expected to be produced by the scatterings of CR proton and helium off the interstellar medium, made by hydrogen and helium at rest. The inclusive cross sections for the possible interactions induce a significant uncertainty in the determination of the antiproton source term and finally on the antiproton flux. 
\\
In  this paper we have determined the requirements on the kinematic parameter space 
that the $p+p\rightarrow \pb+X$ cross section measurements should cover in order 
not to induce uncertainties in the theoretical predictions exceeding the ones inherent in the CR $\pb$ flux data. 
We have assumed that the cross sections could be measured with a few percent accuracy in the relevant regions. 
Our analysis is performed in terms of the $\pb$ source term, which is a proxy of the flux and of its precision level, and is the convolution of the progenitor CR fluxes with the ISM targets and the relevant production cross sections.  
Our results are discussed both in the center-of-mass reference frame, suitable for collider experiments, and in the laboratory frame, as occurring  in the Galaxy. 
To cover all the AMS-02 $\pb$ energy range with the cosmic data precision level, which are now of the order of 5\%, 
one should collect $p+p \rightarrow \pb+X$ cross section data with proton beams from 10 GeV to 6 TeV and pseudorapidity $\eta$ increasing from 2 to nearly 8. 
Alternatively, a full coverage of the CM parameter space should scan $\pT$ from 0.04 to 2 GeV, and $\xR$ from 0.02 to 0.7. 
Similar requirements are found for the $p+\mathrm{He}\rightarrow \pb+X$ and $\mathrm{He} + p\rightarrow \pb+X$  channels. 
These conclusions are not affected by different choices of the cross section parametrization. 
The necessary kinematical coverage is still far from the present collection of data, but it could be fulfilled by fixed target experiments with energies from tens of GeV up to few TeV, 
which are in the reach of CERN accelerators.

\section*{\label{sec::acknowledgments}Acknowledgments}
We would like to thank A. Cuoco, G. Graziani, D. Maurin, G. Passaleva and M.W. Winkler for very useful discussions. 
This work is supported by  the research grants {\sl TAsP (Theoretical Astroparticle Physics)} and {\sl Fermi} funded by the Istituto Nazionale di Fisica Nucleare (INFN). 
M. di Mauro acknowledges support by the NASA Fermi Guest Investigator Program 2014 through the Fermi multiyear Large Program No. 81303 (P.I. E.~Charles).

\bibliography{bibliography}{}
\bibliographystyle{myUnsrt}

\appendix

\section{Useful kinematics }
\label{app::appendix}

\subsection*{Maximal energy of product particles}

Assuming the generic process $a + b \rightarrow c + X$ in the CM frame we
solve for the total energy $E_c$.

\begin{align}
  && m_X^2= p_X^2 &= (p_a + p_b - p_c)^2               \nonumber  \\
  &&              &= s + m_c^2 - 2(p_a+p_b)\cdot p_c   \nonumber  \\
  &&              &= s + m_c^2 - 2 \sqrt{s} E_c        \nonumber  \\
\Rightarrow&& E_c &=   \frac{s+m_c^2-m_X^2}{2\sqrt{s}}
\end{align}
Here $p_i$ are the four-momenta of particle $i$.
In the case of $\pb$ production in $pp$ scattering, we have $m_c=m_p$ 
and, due to baryon number conservation, $m_{X,\mathrm{min}}=3m_p$.
Therefore, the maximal energy allowed for the produce antiproton - which enters in the 
definition of $x_R$ - is:
\begin{align}
  \label{eqn::E_pbar_max}
  E_{\pb, \max} = \frac{s-8m_p^2}{2\sqrt{s}}
\end{align}

\subsection*{Inertial frames and Lorentz transformation}

During the analysis we use two inertial frames. On the one hand, there is the CM frame of the proton-proton or, in the more general case, the nucleon-nucleon scattering. Variables in this system are denoted with a * superscript in the following.
On the other hand, in the LAB frame one of the particles is at rest.  
The LI square of the CM energy is 
\begin{align}
\label{eqn::LT_s}
  s               &= 4 (E^*)^2                    = 2E m + 2 m^2 ,
\end{align}
where $E^*$ is the total energy of the two protons in the CM frame,  $E$ is the incident proton energy in LAB frame,
and $m$ is the proton (nucleon) mass. 
Formally the relation between energy $E$ and momentum $p$ between the two frames is given by the Lorentz transformation
\begin{align}
  E  &= \gamma^*         E^* + \gamma^*\beta^* p^*  \\
  p  &= \gamma^* \beta^* E^* + \gamma^*        p^*.
\end{align}
Here $\beta^*$ is the particle velocity in terms of the speed of light $c$ (we use the convention $c=1$) 
and  $\gamma*=\sqrt{1-(\beta^*)^2}$ is the corresponding Lorentz factor. All of them are linked as follows: 
\begin{align}
  \beta^*         &= p^*/E^*                      &&= \sqrt{\frac{E-m}{E+m}}     &&= \sqrt{\frac{s-4m^2}{s}}  ,  \\
  \gamma^*        &= E^*/m                        &&= \sqrt{\frac{E+m}{2m}}      &&= \frac{\sqrt{s}}{2m_p}    ,  \\
  \gamma^*\beta^* &= \frac{p^*}{m}                &&= \sqrt{\frac{E-m}{2m}}      &&= \frac{\sqrt{s-4m^2}}{2m}  .
\end{align}

\subsection*{Relation of kinetic variables}

Here we give explicitly the relation between the CM frame variables 
$\{\sS, \xR, \pT\}$ and the LAB frame variables $\{T, \Tpbar, \eta\}$. 
From \eqnref{LT_s} we infer
\begin{eqnarray}
  s = 2Tm + 4m^2.
\end{eqnarray}
The transverse momentum is invariant under Lorentz transformation
\begin{eqnarray}
  \pT = p \sin(\theta) =  \sqrt{T^2+2 T m}/\cosh(\eta),
\end{eqnarray}
where we used $\sin(\theta) = 1/\cosh(\eta)$ for $\theta \in [0,\pi]$.

Finally, we get: 
\begin{eqnarray}
  \xR        &=& \frac{E_\pb^*}{E_{\pb, \max}^*} =  \frac{2\sqrt{s}E_\pb}{s-8m_p} \text{, with} \\ \nonumber
  E_\pb^*    &=& \gamma^*         E_\pb - \gamma^*\beta^* {\pL}_\pb \text{ and } \\ \nonumber
  {\pL}_\pb    &=& \cos(\theta) p_\pb = \tanh(\eta) p_\pb.
\end{eqnarray}

\subsection*{Energy-differential and invariant cross section}
\begin{eqnarray}
\label{eqn::energyDifferentialToInvCS_full}
\frac{d\sigma}{d \Tpbar}(T , \Tpbar) 
  &=& \int d\Omega \,\,            \frac{d^3\sigma}{dE_\pb d\Omega} \nonumber \\
  &=& \int\limits_{0}^{2\pi}d\varphi \int\limits_{-1}^{1} d\cos(\theta) \,\, p_\pb^2 \, \frac{dp_\pb}{dE_\pb} \, 
        \frac{d^3\sigma}{p_\pb^2 dp_\pb d\Omega} \nonumber \\
  &=&  2\pi                          \int\limits_{-1}^{1} d\cos(\theta) \,\, p_\pb^2 \, \frac{ E_\pb}{ p_\pb} \, 
        \frac{1}{E_\pb} \sigma_\mathrm{inv}  \nonumber \\
  &=&  2\pi   p_\pb                      \int\limits_{-1}^{1} d\cos(\theta) \,\,  \sigma_\mathrm{inv} \nonumber \\
  &=&  2\pi   p_\pb  \int\limits_{-\infty}^{\infty} d\eta \,\, \frac{1}{\cosh^2(\eta)} \,  \sigma_\mathrm{inv} 
\end{eqnarray}
We used $\cos(\theta) = \tanh(\eta)$ in the last line.

\section{Parameter space covered by $pp$ experiments. Additional plots.}
\label{app::old_data}

\begin{figure*}[t!]
	{  \includegraphics[width=0.45\textwidth]{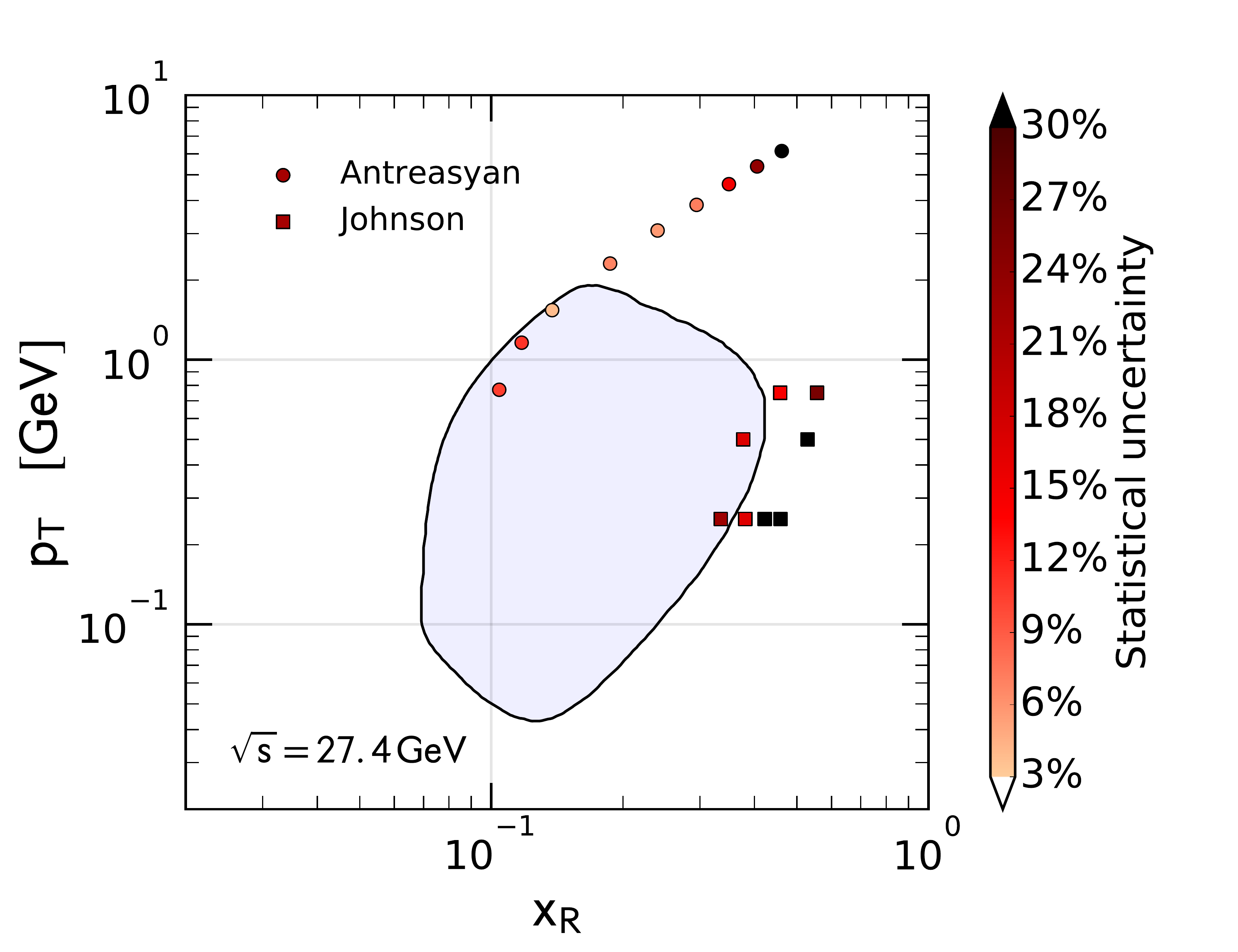}   }
	{  \includegraphics[width=0.45\textwidth]{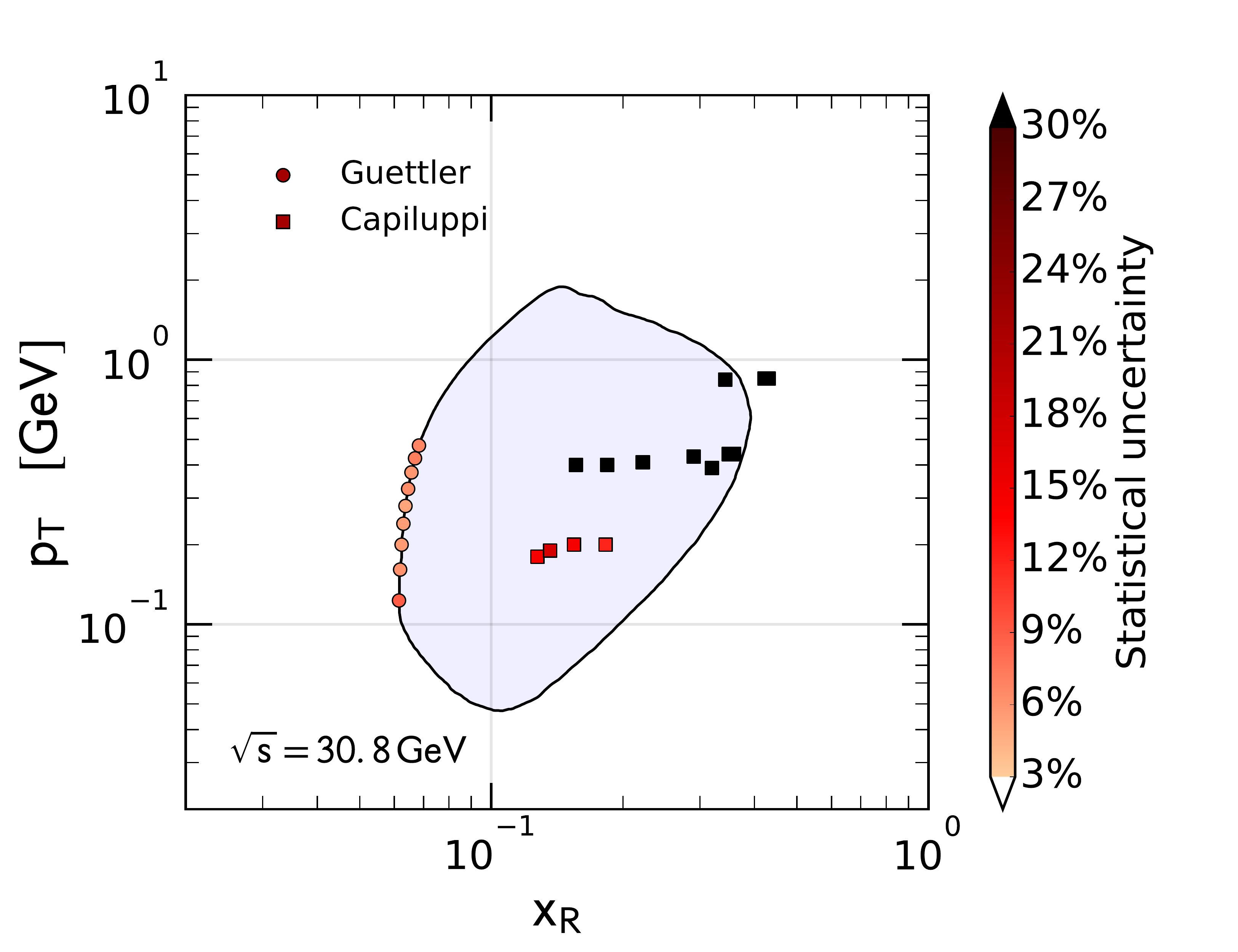}   }\\
	{  \includegraphics[width=0.45\textwidth]{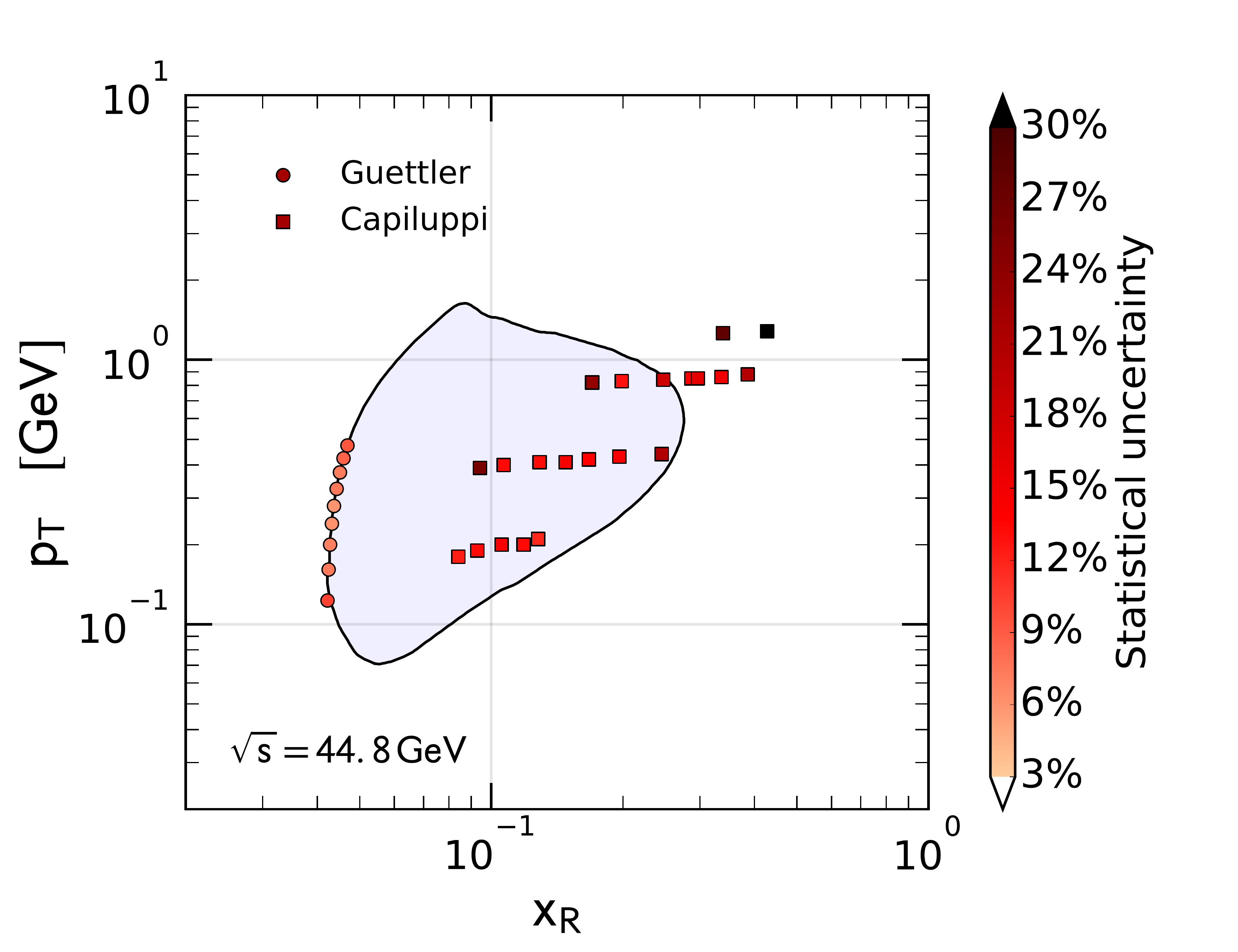}   }
	{  \includegraphics[width=0.45\textwidth]{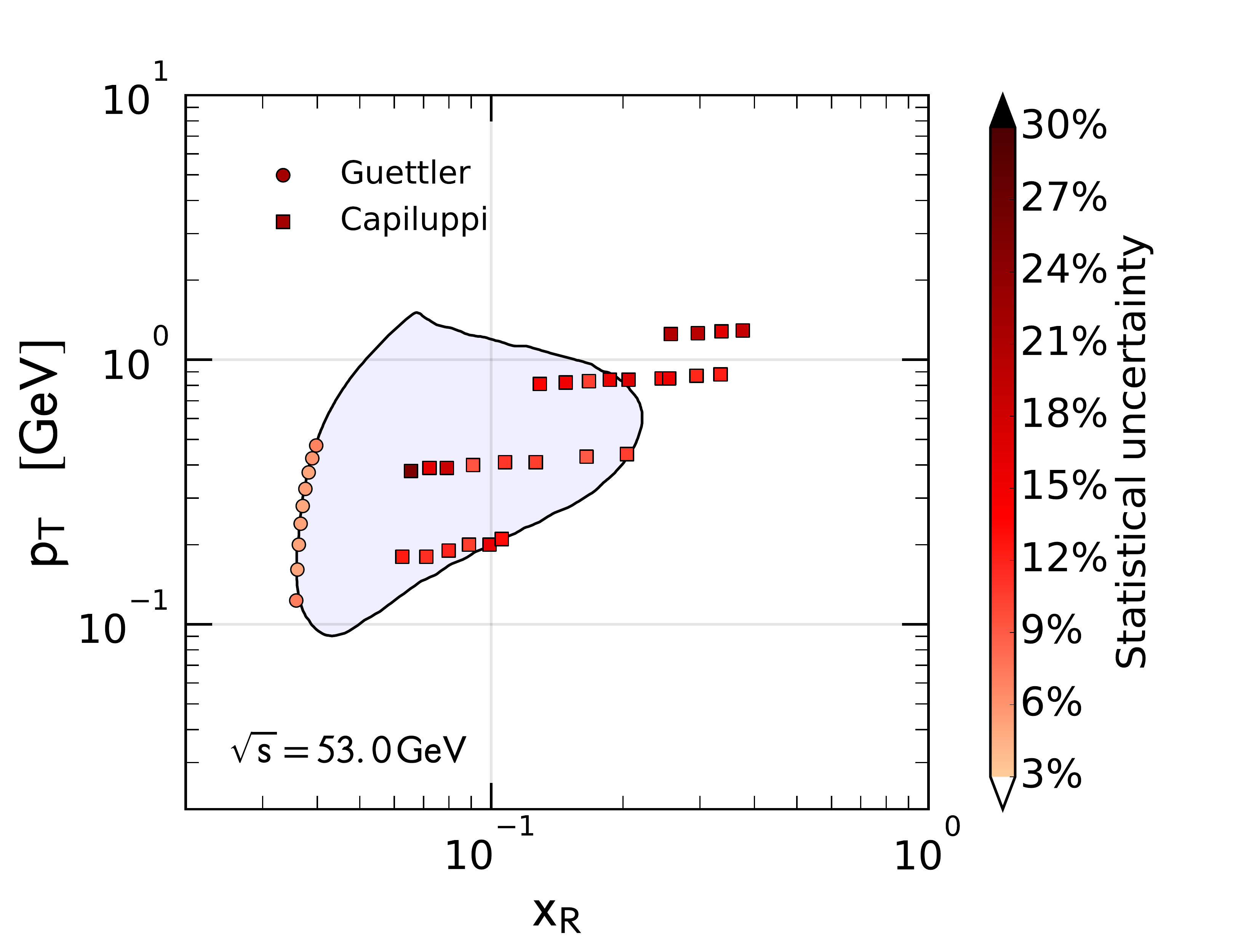}   }\\
	{  \includegraphics[width=0.45\textwidth]{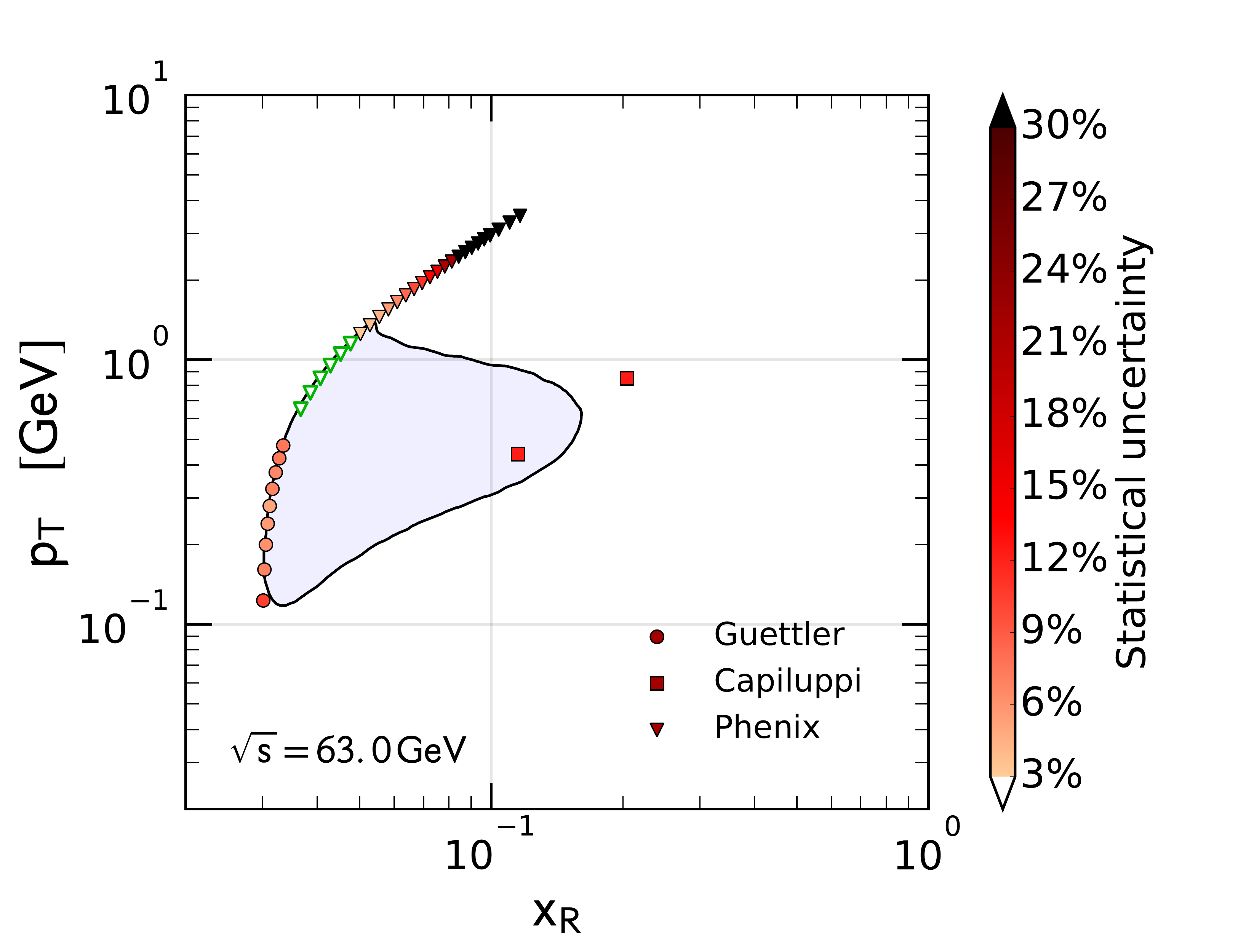}   } 
	\caption{Same as \figref{DataComparison_1}, but for data at increasing $\sqrt{s}.$}
	\label{fig::DataComparison_2}
\end{figure*}

We extend the information already given in \figref{DataComparison_NA49} 
to several  $p+p\rightarrow \pb+X$ cross sections data prior to NA49. 
The data are shown in \figref{DataComparison_1} and \ref{fig::DataComparison_2} 
for increasing $\sqrt{s}$ as functions of $\pT$ and $x_R$. 
It is evident from these figures that the data have large statistical errors  and do not properly cover
the parameter space region required by the AMS-02 $\pb$ data.  
The overall systematic errors  have been omitted here but are reported in \tabref{XS_data}. 
These figures are a simplified visualization of the conclusions already gotten from the analysis 
of this data as performed, i.e.,  in \cite{diMauro:2014_pbarCrossSectionparametrization}.

\begin{table}[h]
  \caption{Summary of cross section measurements and their systematical uncertainties 
            not shown in \figref{DataComparison_1} and \ref{fig::DataComparison_2}.
            Note that \cite{Dekkers:1965zz} and \cite{Capiluppi:1974rt}  do not distinguish between statistical and 
            systematical uncertainties.}
  \label{tab::XS_data}
  \centering
  \begin{tabular}{l c l}
  \hline \hline
  \textbf{Experiment}        & \textbf{Systematic } & \textbf{Ref.}        \\
                             & \textbf{uncertainty} &                      \\ \hline   \hline
  Dekkers et al, CERN 1965          &  -           &  \cite{Dekkers:1965zz}        \\
  Allaby et al, CERN 1970           &  15\%        &  \cite{Allaby:1970jt}         \\
  Capiluppi et al, CERN 1974        &  -           &  \cite{Capiluppi:1974rt}      \\
  Guettler et al, CERN 1976         &  4\%         &  \cite{Guettler:1976ce}       \\
  Johnson et al, FNAL 1978          &  7\%         &  \cite{Johnson:1977qx}        \\
  Antreasyan et al, FNAL 1979       &  20\%        &  \cite{Antreasyan:1978cw}     \\
  Phenix, BNL 2007                  &  9.7\%-11\%  &  \cite{Adare:2011vy}          \\
  BRAHMS, BNL 2008                  &  12\%-19\%   &  \cite{Arsene:2007jd}         \\
  NA49, CERN 2010                   &  3.3\%       &  \cite{NA49_Anticic:2010_ppCollision} \\
  \hline \hline
  \end{tabular}
\end{table}

\begin{figure*}[b!]
	{  \includegraphics[width=0.45\textwidth]{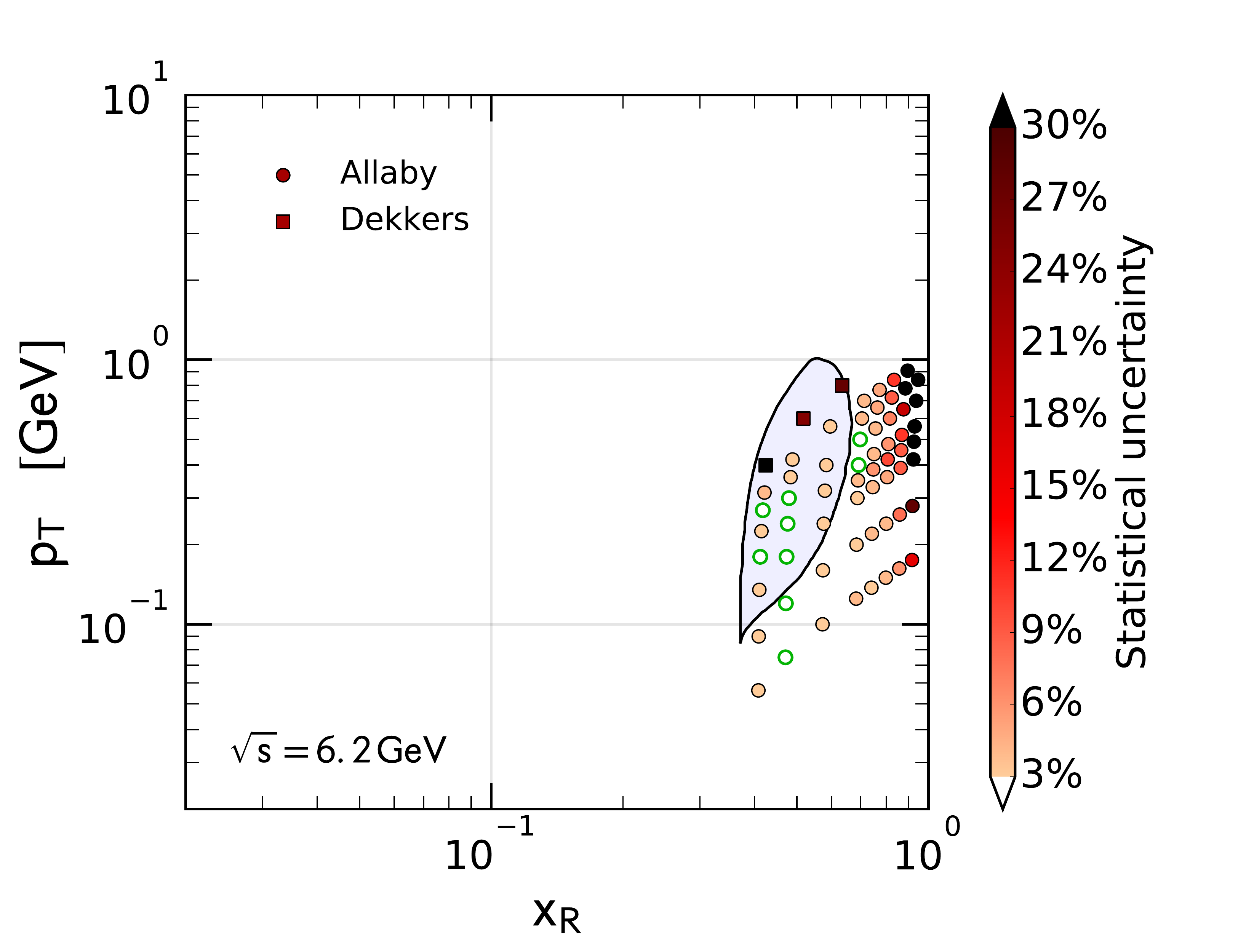}   } 
	{  \includegraphics[width=0.45\textwidth]{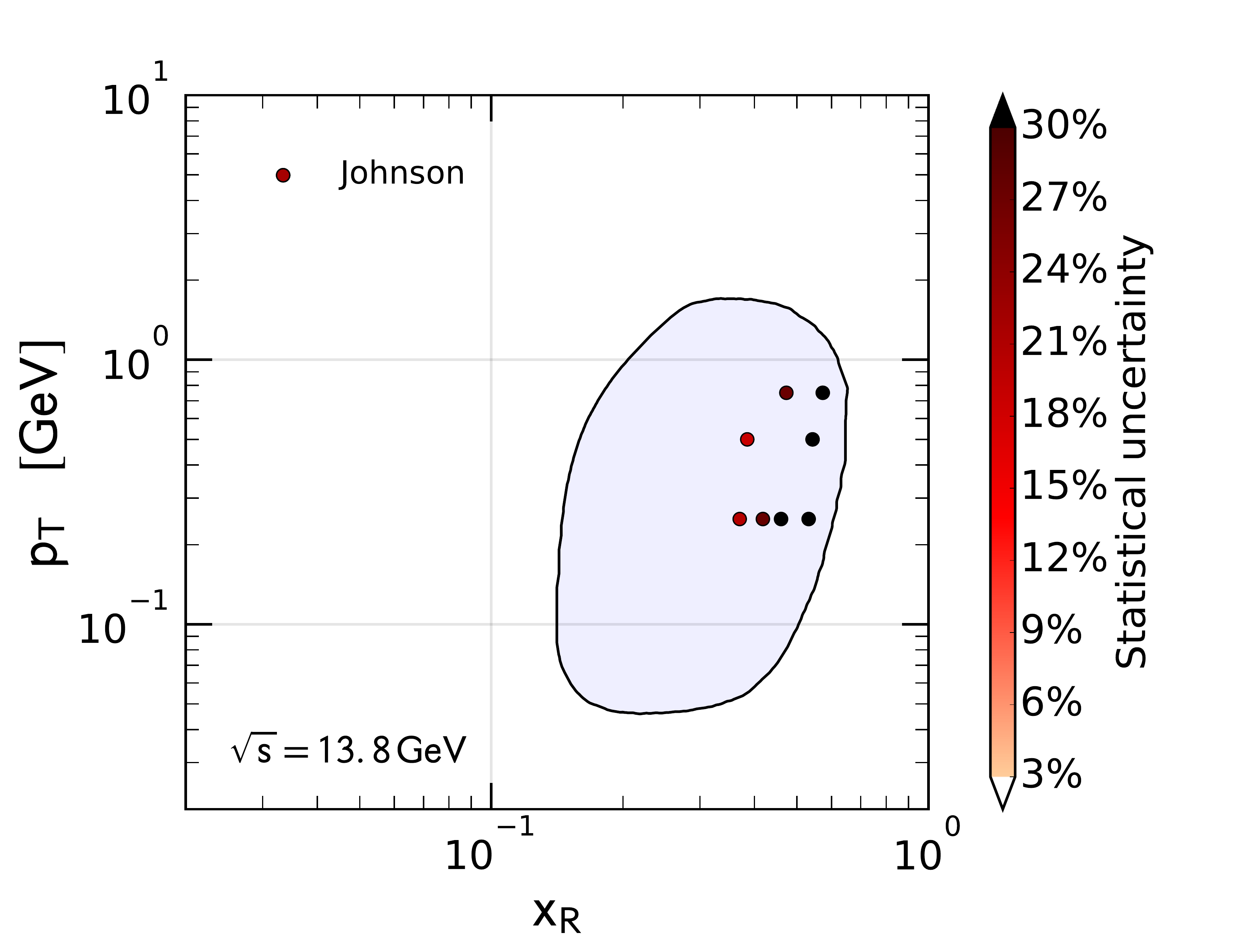}   } \\
	{  \includegraphics[width=0.45\textwidth]{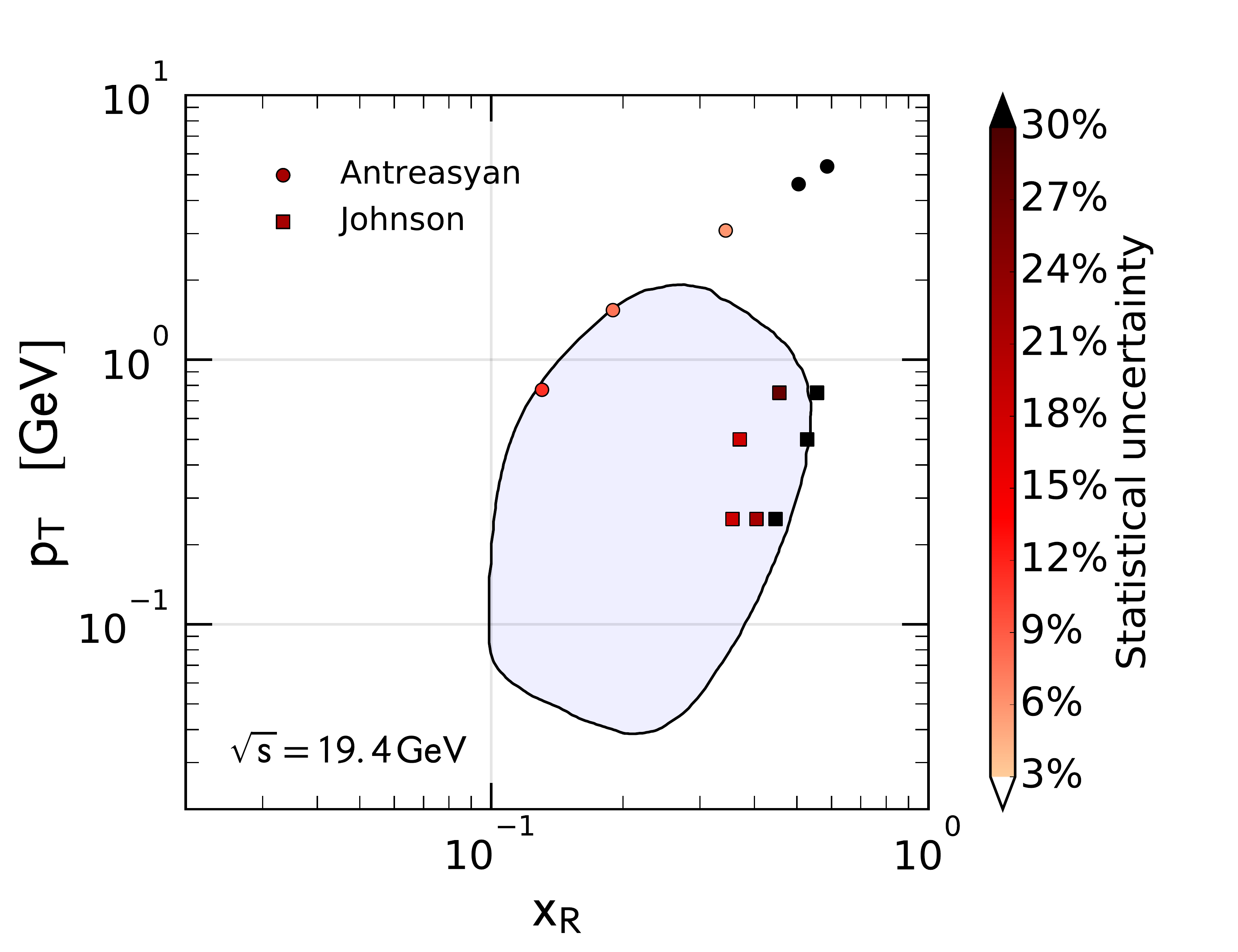}   }
	{  \includegraphics[width=0.45\textwidth]{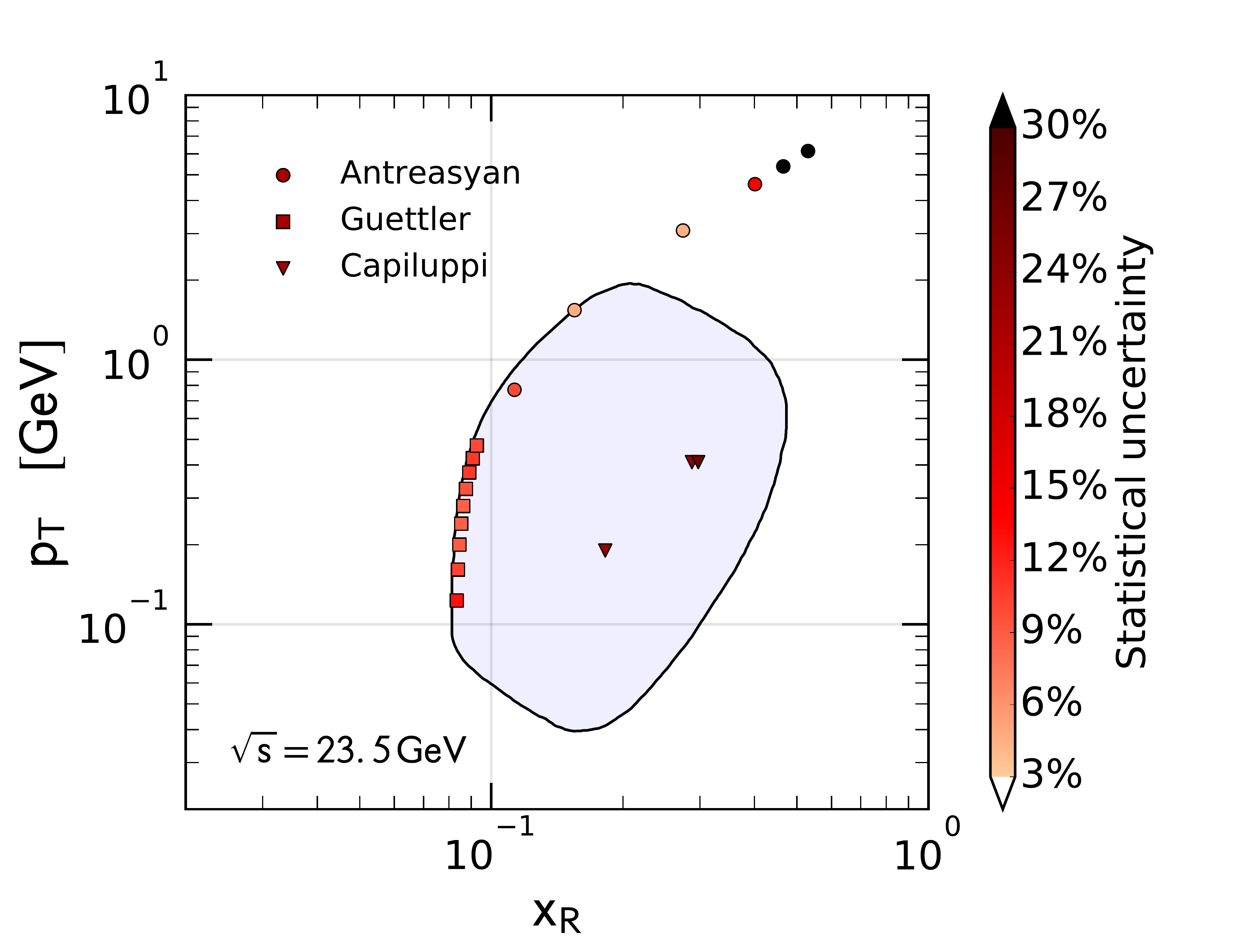}   } 
  \caption{Collection of further experimental data compared to the standard requirement of  
	         3\% accuracy inside the contour and 30\% outside at different values for $\sqrt{s}$.
	         The points with green border fulfill our 3\% requirement.
	         These plots complete the picture presented in \figref{DataComparison_NA49}.
	         We display the statistical uncertainty, while additional systematic uncertainties are 
	         summarized in \tabref{XS_data}.
	         }
	\label{fig::DataComparison_1}
\end{figure*}

\end{document}